%% file: LHCb-PAPER-2024-035.tex
\pdfoutput=1
\documentclass[12pt,a4paper]{article}

\usepackage{multirow}
\usepackage{booktabs}
\usepackage{tikz}
\usepackage{nccmath} 
\usetikzlibrary{decorations.pathreplacing,decorations.pathmorphing,decorations.markings,trees,calc,patterns,snakes,arrows}
\tikzset{
photon/.style={decorate, decoration={snake}, draw=red},
particle/.style={draw=blue, postaction={decorate},decoration={markings,mark=at position .5 with {\arrow[draw=blue]{>}}}},
antiparticle/.style={draw=blue, postaction={decorate},decoration={markings,mark=at position .5 with {\arrow[draw=blue]{<}}}}, 
gluon/.style={decorate, draw=black,decoration={coil,amplitude=4pt, segment length=5pt}}, 
majorana/.style={draw=black, postaction={decorate},decoration={markings,mark=at position .48 with {\arrow[draw=black]{>}},mark=at position .52 with {\arrow[draw=black]{<}}}},
gluonloop/.style={circle, decorate, draw=black, decoration={coil,aspect=1.2,amplitude=2pt, segment length=4pt},minimum height=1.2em},
}
\usepackage{placeins}
\usepackage{ifthen} 
\newboolean{pdflatex}
\setboolean{pdflatex}{true} 

\newboolean{articletitles}
\setboolean{articletitles}{true} 

\newboolean{uprightparticles}
\setboolean{uprightparticles}{false} 

\newboolean{inbibliography}
\setboolean{inbibliography}{false} 

\def\paperauthors{LHCb collaboration} 
\def\paperasciititle{New measurements of Bc decays into single charm final states}
\def\papertitle{New measurements of $\Bcp$ decays into single charm final states}
\def\paperkeywords{{High Energy Physics}, {LHCb}} 
\def\papercopyright{\the\year\ CERN for the benefit of the LHCb collaboration} 
\def\paperlicence{CC BY 4.0 licence}
\def\paperlicenceurl{https://creativecommons.org/licenses/by/4.0/}

\newif\ifEnableSectionTOCLinks
\EnableSectionTOCLinksfalse 

\input{preamble}
\usepackage{longtable} 

\def\RDzK{\ensuremath{R_{\Dz \Kp}}\xspace}
\def\RDstarzK{\ensuremath{R_{\Dstarz \Kp}}\xspace}
\def\RDzpi{\ensuremath{R_{\Dz\pip}}\xspace}
\def\RDzh{\ensuremath{R_{\Dz h^+}}\xspace}
\def\RDKstar{\ensuremath{R_{\Dp\Kstarz}}\xspace}
\def\RDsphi{\ensuremath{R_{\Dsp\phi}}\xspace}

\begin{document}

\renewcommand{\thefootnote}{\fnsymbol{footnote}}
\setcounter{footnote}{1}

\input{title-LHCb-PAPER}


\renewcommand{\thefootnote}{\arabic{footnote}}
\setcounter{footnote}{0}

\cleardoublepage


\pagestyle{plain} 
\setcounter{page}{1}
\pagenumbering{arabic}


\input{body}

\input{acknowledgements}


\addcontentsline{toc}{section}{References}
\bibliographystyle{LHCb}
\bibliography{main,standard,LHCb-PAPER,LHCb-DP}

\newpage
\input{Authorship_LHCb-PAPER-2024-035}

\end{document}

%% file: preamble.tex
\usepackage[top=1in, bottom=1.25in, left=1in, right=1in]{geometry}

\usepackage{microtype}
\usepackage{lineno}  
\usepackage{xspace} 
\usepackage{caption} 

\usepackage{graphicx}  
\usepackage{color}
\usepackage{colortbl}
\graphicspath{{./figs/}} 

\usepackage{amsmath} 
\usepackage{amssymb}
\usepackage{amsfonts}
\usepackage{upgreek} 

\usepackage[normalem]{ulem} 

\newcommand*\patchAmsMathEnvironmentForLineno[1]{%
\expandafter\let\csname old#1\expandafter\endcsname\csname #1\endcsname
\expandafter\let\csname oldend#1\expandafter\endcsname\csname
end#1\endcsname
 \renewenvironment{#1}%
   {\linenomath\csname old#1\endcsname}%
   {\csname oldend#1\endcsname\endlinenomath}%
}
\newcommand*\patchBothAmsMathEnvironmentsForLineno[1]{%
  \patchAmsMathEnvironmentForLineno{#1}%
  \patchAmsMathEnvironmentForLineno{#1*}%
}
\AtBeginDocument{%
\patchBothAmsMathEnvironmentsForLineno{equation}%
\patchBothAmsMathEnvironmentsForLineno{align}%
\patchBothAmsMathEnvironmentsForLineno{flalign}%
\patchBothAmsMathEnvironmentsForLineno{alignat}%
\patchBothAmsMathEnvironmentsForLineno{gather}%
\patchBothAmsMathEnvironmentsForLineno{multline}%
\patchBothAmsMathEnvironmentsForLineno{eqnarray}%
}

\usepackage[pdftex,
            pdfauthor={\paperauthors},
            pdftitle={\paperasciititle},
            pdfkeywords={\paperkeywords}]{hyperref}
\usepackage{hyperxmp}
\hypersetup{
    pdfcopyright={Copyright (C) \papercopyright},
    pdflicenseurl={\paperlicenceurl}
}

\usepackage[colorinlistoftodos,textsize=scriptsize]{todonotes}

\usepackage[bottom,flushmargin,hang,multiple]{footmisc}

\usepackage[all]{hypcap} 

\input{lhcb-symbols-def} 

\hypersetup{
  colorlinks   = true, 
  urlcolor     = blue, 
  linkcolor    = blue, 
  citecolor    = red   
}

\ifEnableSectionTOCLinks
    \usepackage[explicit]{titlesec} 
    
    \let\oldcontentsline\contentsline
    \renewcommand

    \titleformat{\section}{\normalfont\Large\bf}{\hyperlink{tocsection.\thesection}{{\thesection} \parbox[t]{\dimexpr\textwidth-1pc}{#1}}}{1pc}{}

    \titleformat{\subsection}{\normalfont\bf}{\hyperlink{tocsubsection.\thesubsection}{{\thesubsection} \parbox[t]{\dimexpr\textwidth-1pc}{#1}}}{1pc}{}

    \titleformat{name=\section,numberless}[display]{}{}{0pt}{\normalfont\Huge\bfseries #1}
\fi

\usepackage{cite} 
\usepackage{mciteplus}
\makeatletter
\g@addto@macro\bfseries{\boldmath}
\makeatother

%% file: lhcb-symbols-def.tex
\usepackage{xspace} 
\usepackage{upgreek}

\def\lhcb   {\mbox{LHCb}\xspace}

\def\MagUp {\mbox{\em Mag\kern -0.05em Up}\xspace}

\ifthenelse{\boolean{uprightparticles}}%
{

 \def\Ppi         {\ensuremath{\uppi}\xspace}

 \def\PDelta      {\ensuremath{\Delta}\xspace}                 
 \def\PXi         {\ensuremath{\Xi}\xspace}                 
 \def\PLambda     {\ensuremath{\Lambda}\xspace}                 
 \def\PSigma      {\ensuremath{\Sigma}\xspace}                 
 \def\POmega      {\ensuremath{\Omega}\xspace}                 
 \def\PUpsilon    {\ensuremath{\Upsilon}\xspace}
 \let\oldPi\Pi
 \def\PPi         {\ensuremath{\oldPi}\xspace}

 \def\PB      {\ensuremath{\mathrm{B}}\xspace}                 
 \def\PD      {\ensuremath{\mathrm{D}}\xspace}                 

 \def\PK      {\ensuremath{\mathrm{K}}\xspace}                 
 \def\Pb      {\ensuremath{\mathrm{b}}\xspace}                 
 \def\Pc      {\ensuremath{\mathrm{c}}\xspace}

 \def\Ps      {\ensuremath{\mathrm{s}}\xspace}

 \def\thebaroffset{0.0em}
}
{

 \def\Ppi         {\ensuremath{\pi}\xspace}

 \mathchardef\PDelta="7101
 \mathchardef\PXi="7104
 \mathchardef\PLambda="7103
 \mathchardef\PSigma="7106
 \mathchardef\POmega="710A
 \mathchardef\PUpsilon="7107
 \mathchardef\PPi="7105
 \def\PB      {\ensuremath{B}\xspace}                 
 \def\PD      {\ensuremath{D}\xspace}                 

 \def\PK      {\ensuremath{K}\xspace}                 
 \def\Pb      {\ensuremath{b}\xspace}                 
 \def\Pc      {\ensuremath{c}\xspace}

 \def\Ps      {\ensuremath{s}\xspace}

 \def\thebaroffset{0.18em}
}
\newcommand{\offsetoverline}[2][\thebaroffset]{\kern #1\overline{\kern -#1 #2}}%

\makeatletter
\ifcase \@ptsize \relax
  \newcommand{\miniscule}{\@setfontsize\miniscule{4}{5}}
\or
  \newcommand{\miniscule}{\@setfontsize\miniscule{5}{6}}
\or
  \newcommand{\miniscule}{\@setfontsize\miniscule{5}{6}}
\fi
\makeatother

\DeclareRobustCommand{\optbar}[1]{\shortstack{{\miniscule (\rule[.5ex]{1.25em}{.18mm})}
  \\ [-.7ex] $#1$}}

\def\squark    {{\ensuremath{\Ps}}\xspace}

\def\cquark    {{\ensuremath{\Pc}}\xspace}

\def\bquark    {{\ensuremath{\Pb}}\xspace}

\def\pion   {{\ensuremath{\Ppi}}\xspace}
\def\piz    {{\ensuremath{\pion^0}}\xspace}
\def\pip    {{\ensuremath{\pion^+}}\xspace}
\def\pim    {{\ensuremath{\pion^-}}\xspace}

\def\kaon    {{\ensuremath{\PK}}\xspace}
\def\Kbar    {{\ensuremath{\offsetoverline{\PK}}}\xspace}

\def\KorKbar {\kern \thebaroffset\optbar{\kern -\thebaroffset \PK}{}\xspace}

\def\Kp      {{\ensuremath{\kaon^+}}\xspace}
\def\Km      {{\ensuremath{\kaon^-}}\xspace}

\def\Kstarz  {{\ensuremath{\kaon^{*0}}}\xspace}
\def\Kstarzb {{\ensuremath{\Kbar{}^{*0}}}\xspace}

\def\Kstarp  {{\ensuremath{\kaon^{*+}}}\xspace}

\def\Dbar    {{\ensuremath{\offsetoverline{\PD}}}\xspace}
\def\D       {{\ensuremath{\PD}}\xspace}

\def\DorDbar {\kern \thebaroffset\optbar{\kern -\thebaroffset \PD}\xspace}
\def\Dz      {{\ensuremath{\D^0}}\xspace}
\def\Dzb     {{\ensuremath{\Dbar{}^0}}\xspace}
\def\Dp      {{\ensuremath{\D^+}}\xspace}
\def\Dm      {{\ensuremath{\D^-}}\xspace}

\def\DpDm    {\ensuremath{\Dp {\kern -0.16em \Dm}}\xspace}

\def\Dstarz  {{\ensuremath{\D^{*0}}}\xspace}

\def\Dsp     {{\ensuremath{\D^+_\squark}}\xspace}

\def\B       {{\ensuremath{\PB}}\xspace}

\def\BorBbar {\kern \thebaroffset\optbar{\kern -\thebaroffset \PB}\xspace}

\def\Bd      {{\ensuremath{\B^0}}\xspace}

\def\BdorBdbar {\kern \thebaroffset\optbar{\kern -\thebaroffset \Bd}\xspace}
\def\Bu      {{\ensuremath{\B^+}}\xspace}

\def\Bp      {{\ensuremath{\Bu}}\xspace}

\def\Bs      {{\ensuremath{\B^0_\squark}}\xspace}

\def\BsorBsbar {\kern \thebaroffset\optbar{\kern -\thebaroffset \Bs}\xspace}
\def\Bc      {{\ensuremath{\B_\cquark^+}}\xspace}
\def\Bcp     {{\ensuremath{\B_\cquark^+}}\xspace}
\def\Bcm     {{\ensuremath{\B_\cquark^-}}\xspace}

\def\Y#1S{\ensuremath{\PUpsilon{(#1S)}}\xspace}

\def\LorLbar     {\kern \thebaroffset\optbar{\kern -\thebaroffset \PLambda}\xspace}

\def\BF         {{\ensuremath{\mathcal{B}}}\xspace}

\newcommand{\decay}[2]{\mbox{\ensuremath{#1\!\to #2}}\xspace} 

\def\to                 {\ensuremath{\rightarrow}\xspace}

\def\CP                {{\ensuremath{C\!P}}\xspace}

\def\AT#1     {\ensuremath{A_{\mathrm{T}}^{#1}}\xspace}           

\def\C#1      {\ensuremath{\mathcal{C}_{#1}}\xspace}                       
\def\Cp#1     {\ensuremath{\mathcal{C}_{#1}^{'}}\xspace}                    
\def\Ceff#1   {\ensuremath{\mathcal{C}_{#1}^{\mathrm{(eff)}}}\xspace}        
\def\Cpeff#1  {\ensuremath{\mathcal{C}_{#1}^{'\mathrm{(eff)}}}\xspace}       
\def\Ope#1    {\ensuremath{\mathcal{O}_{#1}}\xspace}                       
\def\Opep#1   {\ensuremath{\mathcal{O}_{#1}^{'}}\xspace}                    

\newcommand{\nospaceunit}[1]{\ensuremath{\text{#1}}}       
\newcommand{\aunit}[1]{\ensuremath{\text{\,#1}}}       

\newcommand{\tev}{\aunit{Te\kern -0.1em V}\xspace}
\newcommand{\gev}{\aunit{Ge\kern -0.1em V}\xspace}
\newcommand{\mev}{\aunit{Me\kern -0.1em V}\xspace}
\newcommand{\kev}{\aunit{ke\kern -0.1em V}\xspace}
\newcommand{\ev}{\aunit{e\kern -0.1em V}\xspace}
 
\newcommand{\mevc}{\ensuremath{\aunit{Me\kern -0.1em V\!/}c}\xspace}
\newcommand{\gevc}{\ensuremath{\aunit{Ge\kern -0.1em V\!/}c}\xspace}
\newcommand{\mevcc}{\ensuremath{\aunit{Me\kern -0.1em V\!/}c^2}\xspace}
\newcommand{\gevcc}{\ensuremath{\aunit{Ge\kern -0.1em V\!/}c^2}\xspace}

\def\mum  {\ensuremath{\,\upmu\nospaceunit{m}}\xspace}

\def\fb   {\ensuremath{\aunit{fb}}\xspace}
\def\invfb   {\ensuremath{\fb^{-1}}\xspace}

\def\ps   {\ensuremath{\aunit{ps}}\xspace}

\newcommand{\chisqip}{\ensuremath{\chi^2_{\text{IP}}}\xspace}

\newcommand{\chisqvtx}{\ensuremath{\chi^2_{\text{vtx}}}\xspace}

\def\gsim{{~\raise.15em\hbox{$>$}\kern-.85em
          \lower.35em\hbox{$\sim$}~}\xspace}
\def\lsim{{~\raise.15em\hbox{$<$}\kern-.85em
          \lower.35em\hbox{$\sim$}~}\xspace}

\def\pt         {\ensuremath{p_{\mathrm{T}}}\xspace}

\def\ptot       {\ensuremath{p}\xspace}
\def\et         {\ensuremath{E_{\mathrm{T}}}\xspace}

\def\evtgen     {\mbox{\textsc{EvtGen}}\xspace}

\def\geant      {\mbox{\textsc{Geant4}}\xspace}

\def\photos     {\mbox{\textsc{Photos}}\xspace}

\def\pythia     {\mbox{\textsc{Pythia}}\xspace}

\def\tell1  {TELL1\xspace}
\def\ukl1   {UKL1\xspace}

\newcommand{\ie}{\mbox{\itshape i.e.}\xspace}

\newcommand{\phz}{\phantom{0}}
\newcommand{\lhcborcid}[1]{\href{https://orcid.org/#1}{\hspace*{0.1em}\raisebox{-0.45ex}{\includegraphics[width=1em]{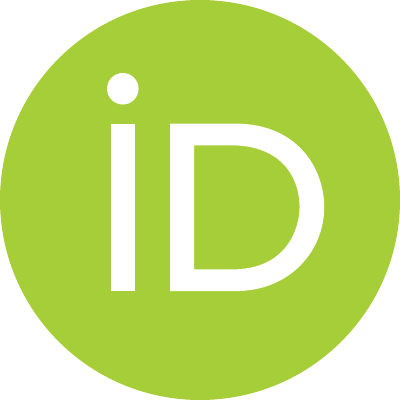}}}}


%% file: title-LHCb-PAPER.tex

\begin{titlepage}
\pagenumbering{roman}

\vspace*{-1.5cm}
\centerline{\large EUROPEAN ORGANIZATION FOR NUCLEAR RESEARCH (CERN)}
\vspace*{1.5cm}
\noindent
\begin{tabular*}{\linewidth}{lc@{\extracolsep{\fill}}r@{\extracolsep{0pt}}}
\ifthenelse{\boolean{pdflatex}}
{\vspace*{-1.5cm}\mbox{\!\!\!\includegraphics[width=.14\textwidth]{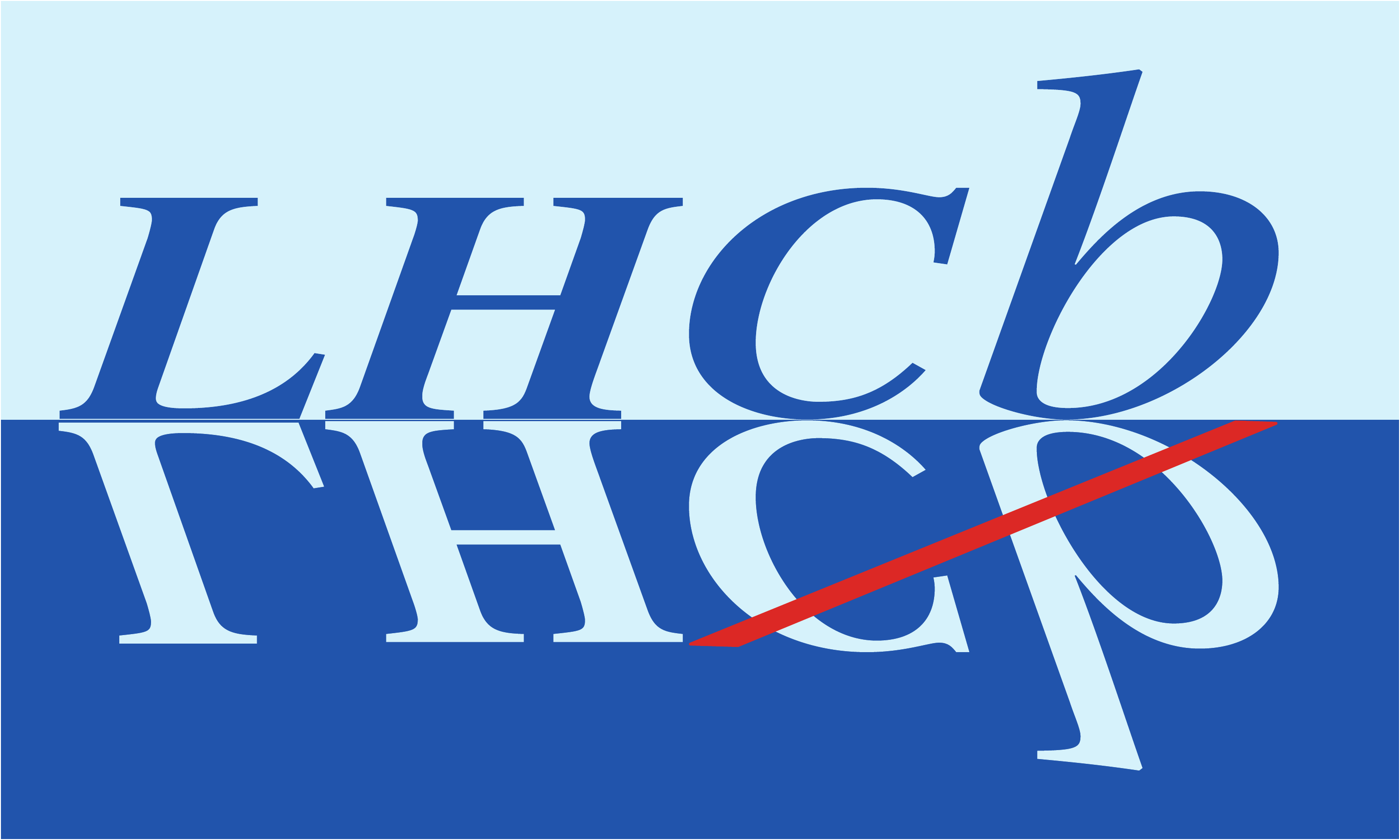}} & &}%
{\vspace*{-1.2cm}\mbox{\!\!\!\includegraphics[width=.12\textwidth]{figs/lhcb-logo.pdf}} & &}%
\\
 & & CERN-EP-2025-088 \\  
 & & LHCb-PAPER-2024-035 \\  
 & & 4 August 2026 \\ 
 & & \\
\end{tabular*}

\vspace*{2.0cm}

{\normalfont\bfseries\boldmath\huge
\begin{center}
  \papertitle 
\end{center}
}


\begin{center}
\paperauthors\footnote{Authors are listed at the end of this paper.}
\end{center}

\vspace{\fill}

\begin{abstract}
  \noindent
  Using proton-proton collision data corresponding to an integrated luminosity of $9 \,\textrm{fb}^{-1}$ collected by the LHCb experiment, searches are performed for $B^+_c$ mesons decaying to a charm and a charmless meson pair. Five products of branching fraction, $\mathcal{B}(B^+_c\!\to DX)$, and fragmentation ratio $f_c\big/f_u$ are reported,
\begin{align*}
R_{D^+ K^{*0}} &= ( 1.42 \pm 0.23 \pm 0.07 \pm 0.11 ) \times 10^{-6}, \\
R_{D^{*0} K^+} &= (1.46 \pm 0.30 \pm 0.11 \pm 0.05 ) \times 10^{-6}, \\  
R_{D^+_s \phi}\ \ \  &= ( 4.0\pm 1.3 \pm 0.2 \pm 0.5) \times 10^{-7 }, \\
R_{D^0 K^+}\ &= ( 9.7 \pm 1.0 \pm 0.4 \pm 0.3 ) \times 10^{-7}, \\
R_{D^0 \pi^+}\ &<\ 1.4 \times 10^{-7}\ \text{at 95\% CL}.
\end{align*}
In each case, the first uncertainty is statistical, the second is systematic and the third includes external uncertainties. The first result is a first observation, the second and third exhibit clear evidence and the fourth improves the precision of previous measurements by a factor two. Additionally, the $C\!P$ asymmetry in $B^+_c\!\to D^0 K^+$ decays is measured and found to be compatible with zero.
\end{abstract}

\vspace*{2.0cm}

\begin{center}
  Submitted to
  JHEP 
\end{center}

\vspace{\fill}

{\footnotesize 
\centerline{\copyright~\papercopyright. \href{\paperlicenceurl}{\paperlicence}.}}
\vspace*{2mm}

\end{titlepage}


\newpage
\setcounter{page}{2}
\mbox{~}
%
%
%
%

%% file: body.tex
\section{Introduction}
\label{sec:Introduction}
 
\noindent
The $\Bcp$ meson is the only ground-state meson composed of two heavy quarks of different flavour.\footnote{Charge conjugation is implied unless otherwise stated.}
Compared to lighter $B^+$ mesons, their formation is suppressed in proton-proton ($pp$) collisions due to the requirement to produce two pairs of heavy quarks. 
Their production at the LHC, relative to that of $B^+$ mesons, is given by the ratio of $b$-quark fragmentation fractions, $f_c/f_u$, which is measured as $(7.26 \pm 1.74)\times10^{-3}$ at $\sqrt{s}=7\tev$ and $(7.56 \pm 1.80)\times10^{-3}$ in $13\tev$ collisions~\cite{LHCb-PAPER-2019-033}. Moreover, the $\Bcp$ lifetime is three times shorter than that of the $B^+$ meson~\cite{LHCb-PAPER-2014-060}, which leads to a reduced selection efficiency because the identification of $\B$ mesons is based heavily on their displacement from the $pp$ collision.

The shorter lifetime is indicative of a larger number of favourable decay paths. 
In contrast to the lighter $B$ mesons, tree-level $W$-emission decays of the $b$ quark account for only $\sim30\%$ of the \Bcp width. 
Around 60\% of the \Bcp decay width is due to a $c$-quark transition, as exemplified by the observed \decay{\Bcp}{B_s^0 \pip} decay~\cite{LHCb-PAPER-2013-044}. 
The remaining 10\% of the decay rate is dominated by weak annihilation, with \decay{b}{s} penguin diagrams accounting for only a small fraction. 
Charmless final states provide an unambiguous signature of weak annihilation in $\Bcp$ mesons but have not yet been observed. 

The most sensitive search to date is that for \decay{\Bcp}{p\bar{p}\pip} decays, which gives an upper limit on the product of the branching fraction and fragmentation fraction ratio $(f_c\big/f_u)\times\mathcal{B}(\decay{\Bcp}{p\bar{p}\pip}) < 3.6\times10^{-8}$ at a 95\% confidence level (CL)~\cite{LHCb-PAPER-2016-001}.
In contrast, the \decay{\Bcp}{\Dz \Kp} decay has been observed by LHCb, and a measurement of \mbox{$(f_c\big/f_u)\times\mathcal{B}(\decay{\Bcp}{\Dz\Kp}) = (9.3^{+2.8}_{-2.5} \pm 0.6) \times 10^{-7}$} was reported~\cite{LHCb-PAPER-2016-058}.  
In the same analysis, no evidence of the \decay{\Bcp}{\Dz \pip} decay was found, which suggests a dominance of weak annihilation over \decay{b}{u} transitions in the observed decay.

\begin{figure}[b]
    \centering
    \includegraphics[width=\linewidth]{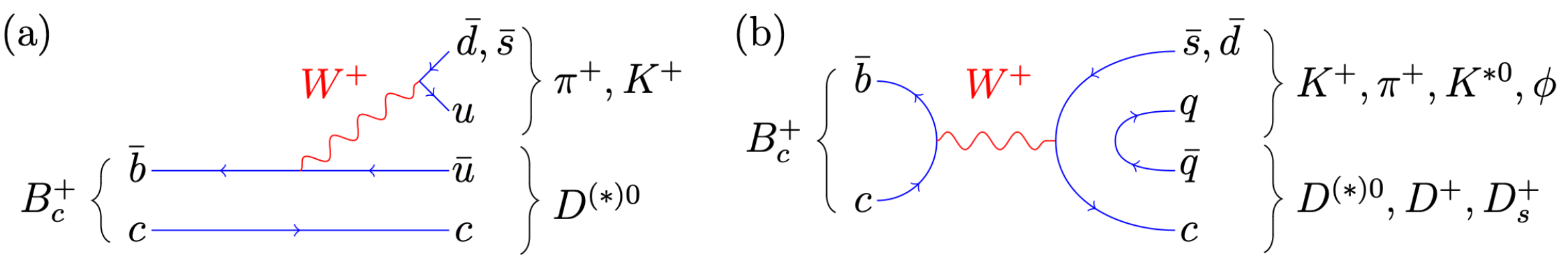}
    \caption{Contributions to $B^+_c\!\to D X$ decays considered here: (a) $W$-emission and (b) weak annihilation. Gluonic penguins are expected to contribute negligibly. The specific final state depends on the flavour of quark pair, $q\overline q$, emerging in the hadronisation.
    }
    \label{fig:DX_diagrams}
\end{figure}

The weak annihilation process in \decay{\Bcp}{\Dz \Kp} decays involves the production of a $u\bar{u}$ quark pair in the final-state hadronisation. 
Assuming weak annihilation is the dominant mechanism, cases where hadronisation produces $d\overline{d}$ or $s\overline{s}$ pairs instead should have a similar rate; see Fig.~\ref{fig:DX_diagrams}. Hence, to further establish the dominance of weak annihilation in single-charm final states, a search for the related decays \decay{\Bcp}{\Dstarz \Kp}, \decay{\Bcp}{\Dp \Kstarz}, and \decay{\Bcp}{\Dsp \phi} is motivated.\footnote{The symbols $\Kstarz$ and $\phi$ refer to the $K^{*}(892)^{0}$ and $\phi(1020)$ states.} These quasi-two-body decays are subsets of the three-body \decay{\Bcp}{Dh^+h^-} modes recently reported~\cite{LHCb-PAPER-2025-028}.

Five observables are reported: \RDzK, \RDzpi, \RDstarzK, \RDKstar, and \RDsphi, measured using $pp$ collision data corresponding to a total integrated luminosity of 9\invfb. 
This includes 3\invfb collected at centre-of-mass energies of $\sqrt{s} = 7$ and $8 \tev$ during Run 1 (2011–2012) and 6\invfb collected at $\sqrt{s} = 13\tev$ during Run 2 (2015–2018). 

The observable $R_{DX}$ is defined as the product of the ratio of the fragmentation fraction into \Bcp and \Bp mesons and the \decay{\Bcp}{DX} branching fraction, where the symbol $D$ represents the charm meson and $X$ represents the charmless meson. 
They are determined from the yield of reconstructed signal candidates that form distinct peaks in the \Bcp invariant-mass distribution.
Measurements are normalised relative to more-abundant $\Bp$ decays that are reconstructed in similar final states, accounting for relative efficiency differences.
Partially reconstructed \decay{\Bcp}{[\Dz \piz]_\Dstarz h^+} and \decay{\Bcp}{[\Dz \gamma]_\Dstarz h^+} decays, where the neutral particles ($\piz,\gamma$) are not reconstructed, also form a characteristic structure in the invariant-mass distribution and are treated as an additional signal channel. 
In order to minimise systematic effects, the normalisation modes are chosen to have the same track multiplicity as the signal modes. 
Table~\ref{tab:Signals} shows a summary of the signal and normalisation modes.

\begin{table}[!bt]
\centering
\captionof{table}{The signal modes with their respective normalisation modes for the measurement of each $R_{DX}$ observable. The $D$ decay, common between signal and normalisation modes, is also listed. The notation 
$[XY]_Z$ indicates a two-body final state $XY$ originating from state $Z$.\label{tab:Signals}}
\begin{footnotesize}
\begin{tabular}{l l l l l}
\toprule
Observable & Signal &\ \  Normalisation& Common $D$ mode\\ 
\midrule
\RDKstar & \decay{\Bcp}{\Dp [\Kp\, \pim\,]_\Kstarz} & \ \ \decay{\Bp}{\Dp[\Kp\, \pim\,]_\Dzb} & \decay{\Dp}{\Km \pip \pip}\\
\RDsphi & \decay{\Bcp}{\Dsp [\Kp \Km]_\phi} & \ \ \decay{\Bp}{\Dsp [\Kp \Km]_\Dzb} & \decay{\Dsp}{\Kp \Km \pip}\\
$R_{\Dstarz\Kp}$ & \decay{\Bcp}{\Dstarz \Kp} & \multirow{3}{*}{$\Bigg\}$\decay{\Bp}{\Dzb\pip}} & \multirow{3}{*}{\decay{\Dz}{\Km\pip}}\\
$R_{\Dz\Kp}$ & \decay{\Bcp}{\Dz \Kp} & & \\
\RDzpi & \decay{\Bcp}{\Dz \pip} &  &\\
\bottomrule
\end{tabular}
\end{footnotesize}
\end{table}

Including candidate yields, $\mathcal{N}$, total efficiencies, $\varepsilon$, and the branching fractions, \BF, of established normalisation decay modes~\cite{PDG2024}, the $R_{DX}$ observables are calculated as follows: 

\begin{footnotesize}
\refstepcounter{equation}
\makeatletter
\renewcommand{\tagform@}[1]{\normalsize(\ignorespaces#1\unskip)} 
\begin{align}
\RDzh &= \frac{\mathcal{N}(\Bcp \to \Dz h^+)}{\mathcal{N}(\decay{\Bp}{\Dzb \pip})} \ \ \,\cdot \mathcal{B}(\decay{\Bp}{\Dzb \pip}) \cdot \frac{\varepsilon(\decay{\Bp}{\Dzb \pip})}{\varepsilon(\Bcp \to \Dz h^+)}, \ \ \ \ h\in\{K,\pi\}, \nonumber \\
\RDstarzK &= \frac{\mathcal{N}(\Bcp \to \Dstarz \Kp)}{\mathcal{N}(\decay{\Bp}{\Dzb \pip})}  \,\cdot \mathcal{B}(\decay{\Bp}{\Dzb \pip}) \cdot \frac{\varepsilon(\decay{\Bp}{\Dzb \pip})}{\varepsilon(\Bcp \to \Dstarz \Kp)}, \ \ \ \ \ \ \ \ \
\nonumber\\
\RDKstar &= \frac{\mathcal{N}(\Bcp \to \Dp K^{*0})}{\mathcal{N}(\decay{\Bp}{\Dp \Dzb})} \cdot \mathcal{B}(\decay{\Bp}{\Dp \Dzb})   \cdot \frac{\varepsilon(\decay{\Bp}{\Dp \Dzb})}{\varepsilon(\Bcp \to \Dp K^{*0})} \cdot \frac{\mathcal{B}(\Dzb \to \Kp \pim)}{\mathcal{B}(K^{*0} \to \Kp \pim)},     \nonumber
\\
\RDsphi &= \frac{\mathcal{N}(\Bcp \to \Dsp \phi)}{\mathcal{N}(\decay{\Bp}{\Dp \Dzb})} \ \,\cdot \mathcal{B}(\decay{\Bp}{\Dp \Dzb})    \cdot \frac{\varepsilon(\decay{\Bp}{\Dp \Dzb})}{\varepsilon(\Bcp \to \Dsp \phi)} \ \,\cdot \frac{\mathcal{B}(\Dzb \to \Kp \Km)}{\mathcal{B}(\phi \to \Kp \Km)}.\tag{\theequation}\label{eq:rdhformula}
\end{align}
\end{footnotesize}
 
Finally, \decay{\Bcp}{\Dz K^+} and \decay{\Bcp}{\Dz \pi^+} decays may exhibit charge-parity (\CP) violation because the $W$-emission amplitude is proportional to the CKM element $V_{ub}^*$, which is known to have a large \CP-violating phase with respect to $V_{cb}^*$, the CKM element present in the weak annihilation diagram. The interference of these two amplitudes can result in a charge asymmetry, though the large difference in magnitude of the \decay{b}{u} and annihilation amplitudes means any interference should be smaller in \decay{\Bcp}{\Dz\Kp} than in \decay{\Bcp}{\Dz \pip} decays, where the \decay{b}{u} transition is more important~\cite{zhang2009branching}.
However, due to the absence of a signal in the \decay{\Bcp}{\Dz \pip} channel, only one \CP asymmetry is reported here,
\begin{equation}
    {A_\CP}(\decay{\Bcp}{\Dz\Kp}) = \frac{\mathcal{N}(\decay{\Bcp}{\Dz\Kp})\, -\, \mathcal{N}(\decay{\Bcm}{\Dzb\Km})}{\mathcal{N}(\decay{\Bcp}{\Dz\Kp})\, +\, \mathcal{N}(\decay{\Bcm}{\Dzb\Km})},\label{eq:cp}
\end{equation}
using yields from separate fits to \Bcp and \Bcm candidates. 

\section{The LHCb detector}
\label{sec:LHCbdetector}

The \lhcb detector~\cite{LHCb-DP-2008-001,LHCb-DP-2014-002} is a single-arm forward
spectrometer covering the \mbox{pseudorapidity} range $2<\eta <5$,
designed for the study of particles containing \bquark or \cquark
quarks. The detector includes a high-precision tracking system
consisting of a silicon-strip vertex detector surrounding the $pp$
interaction region~\cite{LHCb-DP-2014-001}, a large-area silicon-strip detector located
upstream of a dipole magnet with a bending power of about
$4{\mathrm{\,T\,m}}$, and three stations of silicon-strip detectors and straw
drift tubes~\cite{LHCb-DP-2013-003,LHCb-DP-2017-001}
placed downstream of the magnet.
The tracking system measures the momentum, \ptot, of charged particles with
a relative uncertainty ranging from 0.5\% at low momentum to 1.0\% at 200\gevc.
The minimum distance of a track to a primary $pp$ collision vertex (PV), known as the impact parameter (IP), 
is measured with a resolution of $(15+29/\pt)\mum$,
where \pt is the component of the momentum transverse to the beam, in\,\gevc.
Different types of charged hadrons are distinguished using information
from two ring-imaging Cherenkov (RICH) detectors~\cite{LHCb-DP-2012-003}. 
Photons, electrons, and hadrons are identified by a calorimeter system consisting of
scintillating-pad and preshower detectors, an electromagnetic
calorimeter, and a hadronic calorimeter. Muons are identified by a
system composed of alternating layers of iron and multiwire
proportional chambers~\cite{LHCb-DP-2012-002}.
The online event selection is performed by a trigger~\cite{LHCb-DP-2012-004}, 
which consists of a hardware stage based on information from the calorimeter and muon
systems, followed by a software stage that applies a full event
reconstruction.

Simulated samples of signal and normalisation decays are used in the analysis to estimate reconstruction and selection efficiencies and to model the invariant-mass distributions.
In the simulation, $pp$ collisions are generated using
  \pythia~\cite{Sjostrand:2007gs,*Sjostrand:2006za} 
  with a specific \lhcb configuration~\cite{LHCb-PROC-2010-056}. The production of $\Bcp$ decays is modelled using the dedicated \textsc{BcVegPy} generator~\cite{Chang_2004,Chang_2015}. 
  Decays of unstable particles
  are described by \evtgen~\cite{Lange:2001uf}, in which final-state
  radiation is generated using \photos~\cite{davidson2015photos}.
  The interaction of the generated particles with the detector and its response
 is implemented using the \geant
  toolkit~\cite{Allison:2006ve, *Agostinelli:2002hh} as described in
  Ref.~\cite{LHCb-PROC-2011-006}. 
  In some simulated samples, the underlying $pp$ interaction is reused multiple times, each with an independently generated signal decay~\cite{LHCb-DP-2018-004}. For fast simulations, the RapidSim~\cite{Cowan:2016tnm} software package is used.
Additionally, the particle identification (PID) response in the simulation samples is corrected from data using information from calibration samples~\cite{LHCb-PUB-2016-021,LHCb-DP-2018-001}. The applied PID correction is a function of track momentum, pseudorapidity, and the number of tracks in the event.
\section{Selection}
\label{sec:selection}

The selection procedures for all $\Bcp$ signal and $\Bp$ normalisation modes are similar. 
Events are accepted by the hardware trigger if they contain a cluster with large transverse energy, \et, in the hadronic calorimeter~\cite{LHCb-DP-2019-001}. Candidate $B^+_{(c)}$ decays that pass selection are required to have a final-state track associated with this cluster. The selected sample also includes $B^+_{(c)}$ candidates for which the hardware trigger condition was satisfied by tracks that were not part of the signal candidate.
The software trigger requires two, three, or four-track secondary vertices with a significant displacement from any PV. At least one charged track must have $\pt > 1.6\gevc$. A multivariate algorithm~\cite{BBDT} is used to refine the selection of the displaced vertices by requiring that it is consistent with a $b$-hadron decay vertex.

The final selection requires all final-state tracks to be of good quality and to have momentum in the range 5--100\gevc and \pt in the range 0.5--10\gevc; this ensures that tracks are within the kinematic range of the RICH detectors.
Charm-meson candidates are formed from two- or three-track combinations depending on their charge, as listed in Table~\ref{tab:Signals}. Similarly, \Kstarz and $\phi$ candidates are formed from $\Kp\pim$ and $\Kp\Km$ combinations.
The reconstructed masses of \Dz, \Dp, and \Dsp candidates must lie within $\pm25\mevcc$ of their known values~\cite{PDG2024}, whilst those of \Kstarz candidates must be within $\pm100\mevcc$ and $\pm10\mevcc$ of their known values, respectively. 

Candidate decays with a reconstructed \Bcp mass in the interval 5900--6500\mevcc are retained if their decay time is greater than $0.05\ps$. 
Similarly, \Bu candidates with a mass in the interval 5100--5500\mevcc and a decay time greater than $0.20\ps$ are selected. 
The PV with the smallest value of \chisqip is associated with the $B^+_{(c)}$ meson candidate. Here, the \chisqip is the difference in the vertex fit $\chisqvtx$ of the PV reconstructed with and without the particle under consideration.
A kinematic fit~\cite{Hulsbergen:2005pu} is applied to the decay chain where the $B^+_{(c)}$ meson is constrained to originate from the PV, and the invariant mass of the $D$ candidate is constrained to its known value.

For further suppression of random track combinations, a boosted decision tree (BDT) classifier is used~\cite{Roe:2004na}. The classifier is trained using a set of observables that includes track properties such as $p$, \pt, and $\chisqip$, as well as quantities pertaining to the $B^+_{(c)}$ and $D$ meson candidates. 
The latter quantities include the decay time and flight distance, the fit quality of each decay vertex, the radial distance between the decay vertex and the primary vertex, and the angle between the reconstructed momentum vector and the displacement vector between the $B^+_{(c)}$ production and decay vertices. 
An isolation variable that calculates the local imbalance in \pt is also used. It sums the transverse momenta of tracks that lie within a cone around the $B^+_{(c)}$ candidate, excluding the tracks associated with that candidate. True $B^+_{(c)}$ decays tend to be approximately balanced by tracks from an adjacent displaced decay, consistent with $b\overline b$ production.
Lastly, for all tracks in each $B^+_{(c)}$ candidate, PID variables are included in the BDT, except for the companion hadron in $D\Kp$ and $D \pip$ combinations. The PID variable is the difference in the logarithm of the likelihood for the hypothesis of a track being a kaon or a pion, based on information from the RICH detectors.
For the later signal extraction fit, a binary requirement on this PID variable separates \decay{\Bcp}{\Dz\Kp} and \decay{\Bcp}{\Dz\pip} decays.

The BDT classifier is trained using truth-matched simulation samples of the relevant signal and normalisation modes. For the \decay{\Bcp}{[\Km\pip]_\Dz h^+} modes, a background sample of wrong-sign \decay{\Bcp}{[\Kp \pim]_\Dzb \Kp} candidates is chosen. To ensure the background contains only random track combinations, candidates are required to have an invariant mass $m(\Dzb \Kp)>5900 \mevcc$. 
This choice provides a large background sample that is kinematically identical to the signal decay sample.
For the \decay{\Bcp}{\Dp \Kstarz} classifier, wrong-sign \decay{\Bcp}{\Dp \Kstarzb} combinations are used as a background sample.
For the \decay{\Bcp}{\Dsp \phi} classifier, for which a wrong-sign sample cannot be defined, \Bcp candidates in the mass range \mbox{6360--6700\mevcc} are used as a proxy for the background. 
This sample overlaps with the final sample used for yield determination, so to avoid bias, the dataset is randomly divided into two subsamples, and parallel BDT training procedures are followed. Each of the resulting classifiers is applied only to candidates of the other subsample.
For each normalisation mode, the BDT is trained using a background sample of candidates in the high invariant-mass sideband, $m(\Bp)>5900 \mevcc$ (vetoing the \Bcp region).

The BDT selection criterion for each \Bcp mode is optimised by maximising the figure of merit, $S/\sqrt{S+B}$, as a function of the BDT classifier output. Here, the expected signal yield $S$ is compared to the estimated number of background events, $B$, within a $3\sigma$ window around the known $\Bcp$ mass, where the peak width, $\sigma$, is obtained from simulation.
The value of $S$ is estimated by calculating $\mathcal{N}(\decay{\Bcp}{DX})$ from a rearrangement of Eq.~\ref{eq:rdhformula} with the following assumptions: \RDzK takes the value found in Ref.~\cite{LHCb-PAPER-2016-058}; $\RDzpi = 0.05 \RDzK$ accounts for Cabibbo suppression; $\RDKstar = \RDzK$ and $\RDsphi = 0.5\RDKstar$, where the factor 0.5 is a proxy for the reduced likelihood of $s\overline{s}$ gluon splitting in the final state.
The value of $B$ is estimated by counting the background in the relevant samples. For the \decay{\Bcp}{\Dz h^+} channel, the number of wrong-sign \decay{\Bp}{\Dzb \Kp} candidates within a $3\sigma$ window is used. For the \decay{\Bcp}{\Dsp \phi} decay, the value of $B$ is determined by fitting the upper mass sideband region with an exponential function and extrapolating into the signal region.
For the normalisation modes, the BDT selection is optimised by minimising the relative fit uncertainty on the measured normalisation yield. 
In less than 3\% of events, multiple candidates survive all selection, and in these cases, one candidate is selected arbitrarily.

\section{Invariant-mass fits}
\label{sec:massfits}
Signal and normalisation yields are extracted from maximum-likelihood fits to invariant-mass distributions of \Bcp and \Bu candidates. 
All fits are unbinned except for the $\decay{\Bp}{\Dzb \pip}$ normalisation mode, where a binned fit is more practical given the large statistics.
For each mode, the total probability density function (PDF) is built from signal and background components. 

In the normalisation modes, the \Bu peak is modelled by the sum of two Crystal Ball functions~\cite{Skwarnicki:1986xj}, which are composed of a Gaussian core and asymmetric power-law tails that describe non-Gaussian resolution effects. 
The two Gaussian functions share the same mean and width and are left free to vary in the fit to the data. The relative proportion of the two parts varies freely, whilst the tail parameters are fixed from simulation. 
Misidentified backgrounds from \decay{\Bp}{\Dsp\Dzb} decays in the \decay{\Bp}{\Dp \Dzb} sample and \decay{\Bp}{\Dzb \Kp} decays in the $\decay{\Bp}{\Dzb \pip}$ sample are taken into account. Their invariant-mass shapes are derived from simulation, described as a sum of two Crystal Ball functions with fixed parameters. The single-charm, \decay{\Bp}{\Kp\Km\pip\Dzb} background in the \decay{\Bp}{\Dsp\Dzb} sample is defined by performing ancillary fits to the lower and higher $\Dsp$-mass sidebands and interpolating the shape and yield into the signal region. 
For all modes, the combinatorial background is described by an exponential function with the slope freely varying in each fit.
The normalisation yields obtained from these fits are:  \mbox{$\mathcal{N}(\decay{\Bp}{\Dzb \pip}) = 1\,315\,300 \pm 1\,100$}; \mbox{$\mathcal{N}(\decay{\Bp}{\Dp \Dzb}) = 2\,446 \pm  59$}; and \mbox{$\mathcal{N}(\decay{\Bp}{\Dp \Dzb}) = 5\,572 \pm  80$}. The contribution from the Run~2 sample is about 80\% of these yields.

For the $\Bcp$ signal modes, unbinned maximum-likelihood fits are applied to the invariant-mass distributions of \Bcp candidates over the interval \mbox{5900--6500~\mevcc}. 
Individual fits are performed on the \decay{\Bcp}{\Dp \Kstarz} and \decay{\Bcp}{\Dsp \phi} samples, but the \decay{\Bcp}{\Dz\Kp} and \decay{\Bcp}{\Dz\pip} samples are fitted simultaneously so that the misidentified crossfeed yield can be determined dynamically.

In each case, the signal is modelled by the sum of two Crystal Ball functions with a common peak position and width, but the tail parameters and relative contributions are fixed to the values found in the simulation. For the \decay{\Bcp}{\Dz h^+} samples, both the peak position and width vary freely. From this, the \Bcp mass is measured to be $6274.2 \pm 1.8 \mevcc$, consistent with the world average~\cite{PDG2024}. The \decay{\Bcp}{\Dz \pip} mode shares the same peak position as \decay{\Bcp}{\Dz \Kp}, though the width is slightly enlarged by an appropriate ratio taken from simulation. For the \decay{\Bcp}{\Dp \Kstarz} and \decay{\Bcp}{\Dsp \phi} modes, due to the low number of events, the peak position is fixed to the known $\Bcp$ mass value~\cite{PDG2024}, and the width is fixed to the value found in simulation, corrected by the ratio of widths obtained in fits to the normalisation mode data and simulation. 
In the \decay{\Bcp}{\Dz h^+} fit, the misidentified crossfeed components are modelled using the sum of two Crystal Ball functions, with all shape parameters fixed to values obtained from simulation.

Partially reconstructed \Bcp decays, where either a photon or $\piz$ meson is not included in the reconstruction, appear at invariant masses lower than the \Bcp mass. These contributions are modelled using parametric PDFs~\cite{LHCb-PAPER-2020-036} composed of a parabola with positive (negative) curvature, motivated by the $0^-(1^-)$ spin-parity of the missing particle,  convolved with a Gaussian resolution function. The mass of the missing particle defines the kinematic endpoints of these distributions, and all shape parameters are fixed to the values found in simulation.
In the \decay{\Bcp}{\Dz h^+} fits, the low-mass region contains \decay{\Bcp}{[\Dz\piz]_\Dstarz h^+}, \decay{\Bcp}{[\Dz\gamma]_\Dstarz h^+}, and \decay{\Bcp}{\Dz[\Kp\piz]_\Kstarp}  
decays where the $\piz$ and $\gamma$ are not reconstructed.
In the low-mass regions of the \decay{\Bcp}{\Dp \Kstarz} and \decay{\Bcp}{\Dsp \phi} distributions, contributions involving \decay{\D^{*+}_{(s)}}{D^+_{(s)}\piz} and \decay{\D^{*+}_{(s)}}{D^+_{(s)}\gamma} decays are present. 
As with the normalisation modes, combinatorial background PDFs are included in all signal fits modelled by an exponential function with varying slope and yield. 

\begin{figure}[tb]
    \centering
    \includegraphics[width = 0.49\linewidth]{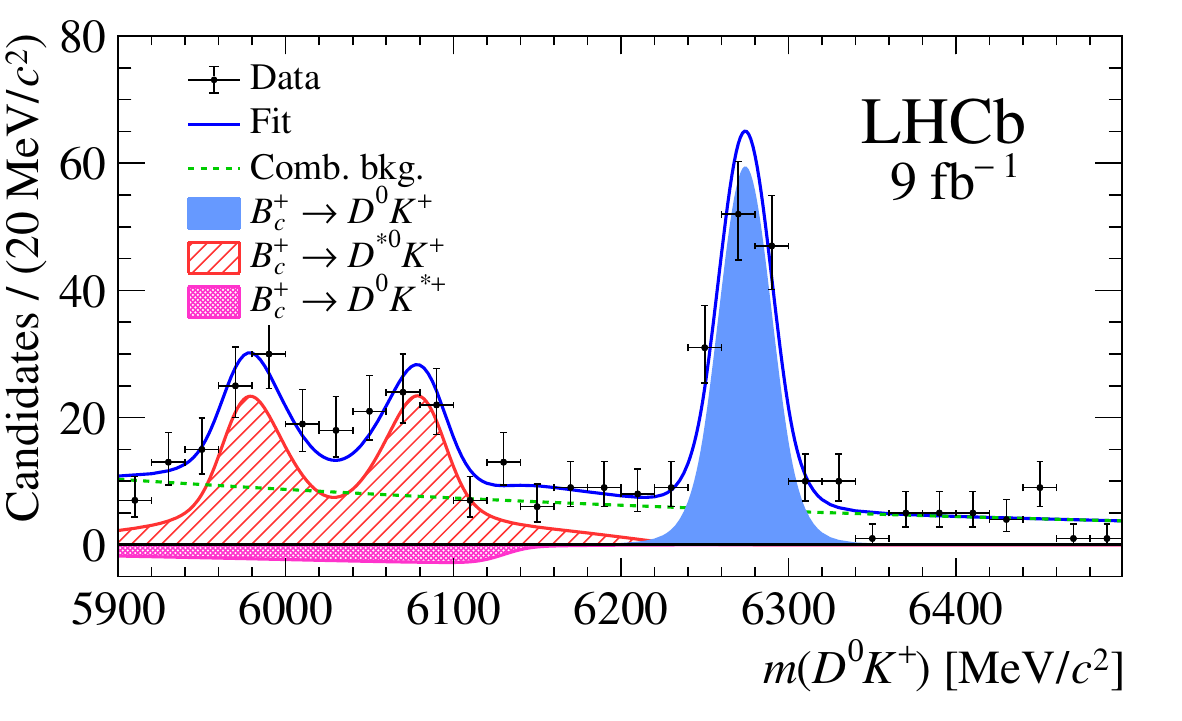}
    \includegraphics[width = 0.49\linewidth]{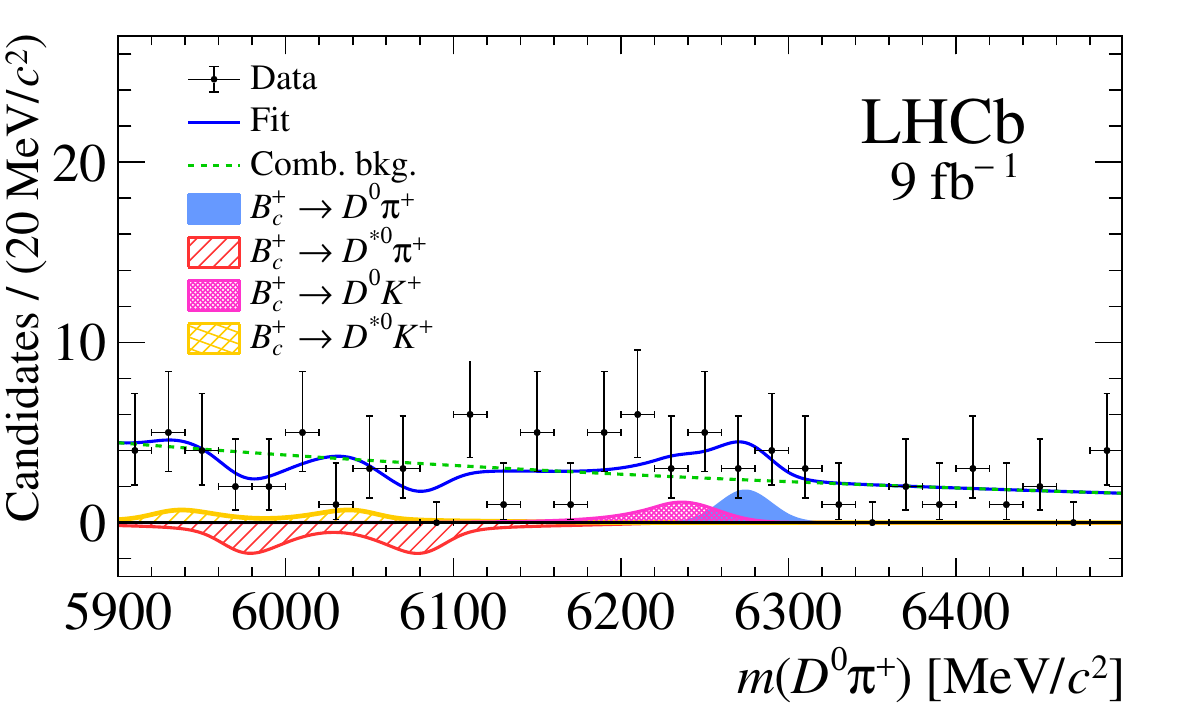}
    \caption{Invariant-mass distribution of (left) $B^+_c\!\to D^0 K^+$ and (right) $B^+_c\!\to D^0 \pi^+$ candidates with the result of the simultaneous fit.
}
    \label{fig:fit1}
\end{figure}

\begin{figure}[tb]
    \centering
    \includegraphics[width = 0.49\linewidth]{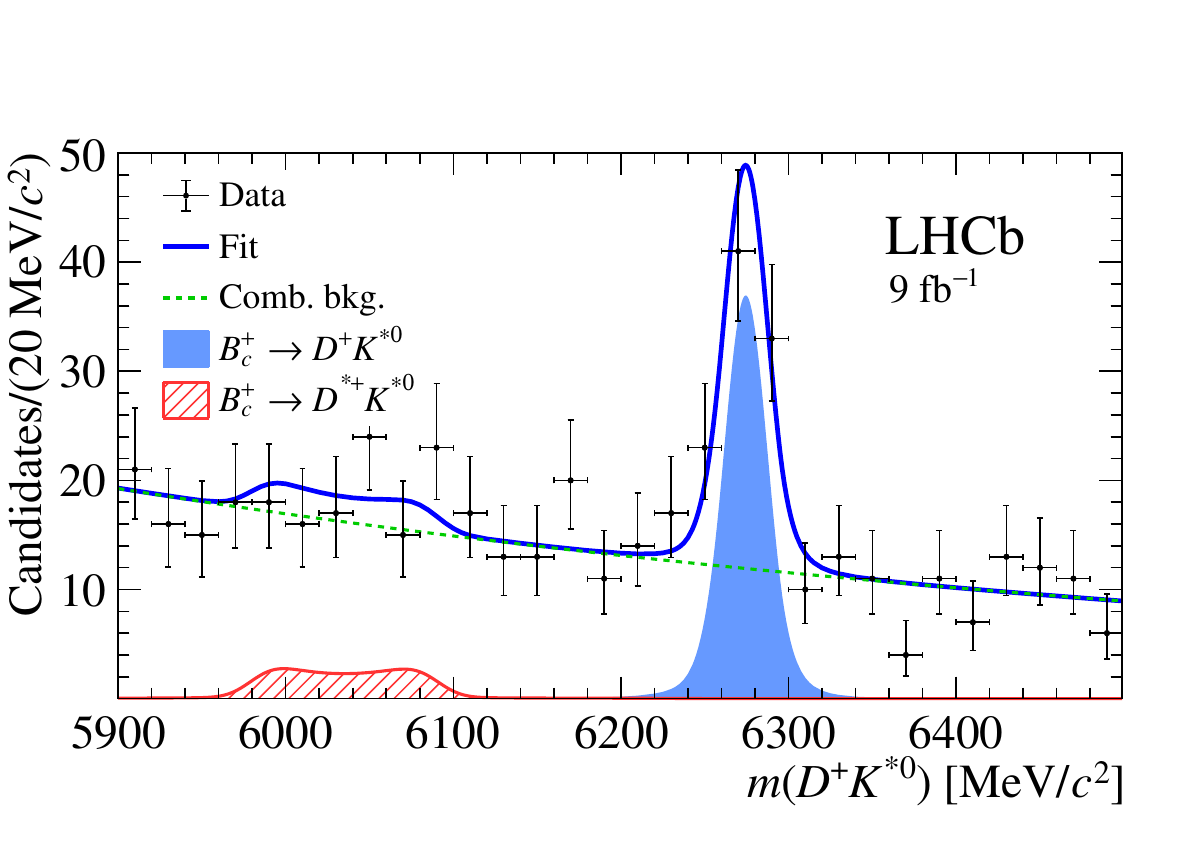}
    \includegraphics[width = 0.49\linewidth]{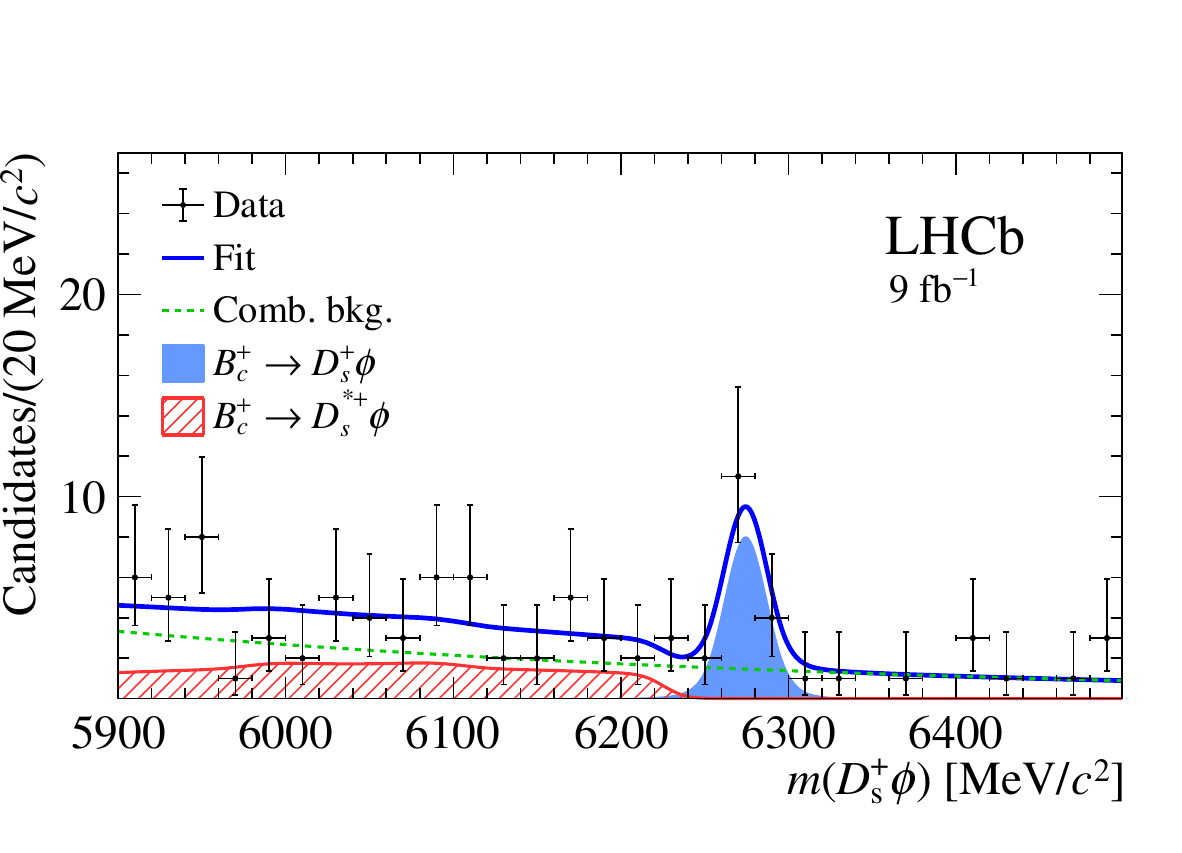}
    \caption{Invariant-mass distribution of (left) $B^+_c\!\to D^+ K^{*0}$ and (right) $B^+_c\!\to D^+_s \phi$ candidates with the result of the fit.}
    \label{fig:fit2}
\end{figure}

The fits to the data are shown in Figs.~\ref{fig:fit1} and \ref{fig:fit2}, and the corresponding signal yields are given in Table~\ref{tab:yields_sig}. 
In the \decay{\Bcp}{\Dz\pip} fit, the total PDF shows dips in the low-mass region that are attributed to the \decay{\Bcp}{D^{*0} \pip} component, finding a negative yield that is compatible with zero.

The significance of each $\Bcp$ signal is determined via a likelihood ratio, taking into account the systematic uncertainty associated with the fit model. The resulting significances are determined from Wilks' theorem~\cite{Wilks:1938dza} and are included in Table~\ref{tab:yields_sig}. 
The \decay{\Bcp}{\Dp \Kstarz} mode is observed for the first time, and clear evidence is seen for \mbox{\decay{\Bcp}{\Dstarz \Kp}} and \mbox{\decay{\Bcp}{\Dsp \phi}} decays.

\begin{table}[tb]
    \centering
    \caption{Signal yields and significances for each $\Bcp$ decay mode. 
    }
    \begin{tabular}{lcc}
    \toprule
     Decay    & Yield & Significance \\
     \midrule
     \decay{\Bcp}{\Dz \Kp} & $129 \pm 14$ & $ 14.5\,\sigma$ \\
     \decay{\Bcp}{\Dp \Kstarz} & \phz$64 \pm 11$ & \phz$8.1\,\sigma$ \\
     \decay{\Bcp}{\Dstarz \Kp} & $152 \pm 32$ & \phz$4.9\,\sigma$ \\
     \decay{\Bcp}{\Dsp \phi} & $14 \pm 4$ & \phz$4.3\,\sigma$ \\
     \decay{\Bcp}{\Dz \pip}    & \phz$4 \pm 4$ & \phz$1.1\,\sigma$ \\
     \bottomrule
    \end{tabular}
    \label{tab:yields_sig}
\end{table}

A measurement of the direct \CP violation in the \decay{\Bcp}{\Dz \Kp} decay is made by performing fits on the final samples split by charge. The same fit model is used in a simultaneous fit to the \Bcp and \Bcm samples.
 The \CP asymmetry of Eq.~\ref{eq:cp} is measured to be $A_\CP(\decay{\Bcp}{\Dz \Kp})= 0.07 \pm 0.10\pm 0.03$, where the first uncertainty is statistical and the second covers systematic effects related to the fit model and the \Bcp production asymmetry\cite{LHCb-PAPER-2019-033}. This result is consistent with zero (no \CP asymmetry). Detection asymmetries play a negligible role because the two kaons in the final state are of opposite charge and have nearly identical momentum spectra.

\section{Efficiencies}
\label{sec:efficiency}

To form the $R_{DX}$ observables of Eq.~\ref{eq:rdhformula}, the selection efficiencies of the signal and normalisation modes are determined from simulated samples after correcting for data-simulation differences. 
Each efficiency is the product of detector acceptance, trigger, reconstruction, and offline selection effects. 
The treatment of the trigger efficiency is most critical because events that fail the trigger are not recorded. Instead, the analysis uses an established data-driven method~\cite{LHCb-PUB-2014-039} that determines efficiencies without knowing the number of events that failed the trigger. This method is applied to the abundant \decay{\Bp}{\Dzb\pip} normalisation mode across a 2D binning of \pt of the \Bp meson and the maximum \pt amongst the final state hadrons, and does this in both data and simulation, after final selection. 
Then, for all signal and normalisation modes, the proportion of events inside each bin is calculated, and an effective trigger efficiency is deduced for both data and simulation. The ratio of these numbers is the trigger correction applied to the prediction from the simulation.

After corrections, all efficiencies are evaluated independently for each year and magnet polarity, then combined using weights proportional to the integrated luminosity of each dataset.  
The resulting efficiency ratios are
\begin{align*}
\frac{\varepsilon(\decay{\Bp}{\Dzb \pip})}{\varepsilon(\decay{\Bcp}{\Dz \Kp})} &= 2.12 \pm 0.08, & \frac{\varepsilon(\decay{\Bp}{\Dzb \pip})}{\varepsilon(\decay{\Bcp}{\Dz \pip})} &= 3.28\pm 0.20, \nonumber\\
\frac{\varepsilon(\decay{\Bp}{\Dp \Dzb})}{\varepsilon(\decay{\Bcp}{\Dp K^{*0})}} &= 2.40 \pm 0.09, & \frac{\varepsilon(\decay{\Bp}{\Dp \Dzb})}{\varepsilon(\decay{\Bcp}{\Dsp \phi)}} &=  2.25 \pm 0.09,\nonumber\\
\frac{\varepsilon(\decay{\Bp}{\Dzb \pip})}{\varepsilon(\decay{\Bcp}{D^{*0} \Kp)}} &= 2.69 \pm 0.20,
\end{align*}
where the uncertainties are the related systematic uncertainties, as described in Sec.~\ref{sec:systematics}.

These efficiency ratios are far from unity because the lifetime of the \Bcp meson is 3.2 times shorter than that of the \Bp meson. Hence, the displaced-vertex requirements are comparatively less performant, reducing the trigger and selection efficiencies. This is partially compensated for by the heaviness of the \Bcp meson, which leads to a harder momentum spectrum for the final-state tracks, thereby improving the trigger performance. The efficiency ratio for the measurement of \RDzpi is markedly larger due to the strict \Bcp selection criteria applied to select the small \decay{\Bcp}{\Dz \pip} signal. For the \decay{\Bcp}{\Dp \Kstarz} and \decay{\Bcp}{\Dsp \phi} signal modes, the decay products have a softer \pt spectrum compared to their normalisation modes, which adversely affects their efficiency.

\section{Systematic uncertainties}
\label{sec:systematics}

Due to the use of ratios in the $R_{DX}$ observables, many sources of systematic uncertainty cancel or are significantly reduced. Apart from the contribution of external branching fractions and a small uncertainty from the choice of fit model, all the systematic uncertainties are related to the evaluation of the efficiencies. 
These are quantified in Table~\ref{tab:allsyst} and described below.

The uncertainty related to the choice of the PDFs in the invariant-mass fit model is evaluated by refitting the data many times with PDFs for which the various fixed parameters are varied together within their uncertainty, as defined in the fit to the simulation samples. The width of the resulting distribution of $R_{DX}$ values is taken as a systematic uncertainty of the principal result.

The finite number of simulated events limits the precision with which any simulation-based quantity can be determined. An uncorrelated variation of signal and normalisation mode efficiencies, based on the statistical uncertainty of the relevant simulation samples, provides an uncertainty on the efficiency ratios and thereby each $R_{DX}$ measurement.

As described in Sec.~\ref{sec:efficiency}, the trigger efficiency is calculated using the $\decay{\Bp}{\Dzb \pip}$ normalisation mode data, applied in bins of the $B^+_{(c)}$ candidate's \pt and the maximum \pt among the final-state tracks.
Though this method is data-driven, it assumes that any mismodelling of the decay kinematics in the simulation is the same in the signal and normalisation samples. To assign an associated systematic error, the double ratio of data and simulation efficiencies for both \Bp and \Bcp is compatible with unity, and the difference is propagated as a systematic uncertainty on the $R_{DX}$ observables. The \RDstarzK measurement carries the largest uncertainty related to the trigger because the weighting procedure works least well for \decay{\Bcp}{\Dstarz\Kp} decays. This is due to the $\pi^0,\gamma$ from the \Dstarz decay not being included in the reconstruction, so the \Bcp momentum is far from the true momentum, and events are assigned to slightly incorrect bins across the 2D efficiency maps. 

The reweighting of the $\decay{\Bp}{\Dzb \pip}$ normalisation mode sample is also used to verify offline selection efficiency ratios. This difference in the BDT response for data and simulation is similar for normalisation mode and the signal decay kinematics (\ie, after reweighting). The discrepancy in the BDT response differs at the level of 1\% for the optimised selection value, and thus a systematic uncertainty of 1\% is assigned across all signal modes.

The efficiency ratio averages over LHCb datasets (year and polarity). To account for small discrepancies in the subsets processed in this work, a small luminosity uncertainty is assigned.
Finally, the total systematic uncertainty is the sum in quadrature of the individual contributions. The published uncertainties on the measured branching fraction values~\cite{PDG2024} used to calculate $R_{DX}$ are included as a separate uncertainty.

\begin{table}[tb]
    \centering
    \caption{Sources and absolute values of the measurement uncertainties.}
    \resizebox{1\columnwidth}{!}{
    \begin{tabular}{lccccc}
    \toprule
        &$\Bcp \to \Dz \Kp$ &$\Bcp \to D^{*0} \Kp$&$\Bcp \to \Dz \pip$ & $\Bcp \to \Dp K^{*0}$ & $\Bcp \to \Dsp \phi$ \\
     & [$\times 10^{-8}$] & [$\times 10^{-8}$] & [$\times 10^{-8}$] & [$\times 10^{-8}$] & [$\times 10^{-8}$] \\
     \midrule
     Statistical uncertainty & 10.4\phz & 30.4\phz & 4.6 & 23.3\phz & 12.9\phz \\
     \midrule
     Total systematic uncertainty &  3.6	& 11.1~	& 0.4	& 6.9	& 1.8 \\
     \midrule
     ~~~~Model parametrisation & 0.3 &	2.2 &	0.3 &	4.3 &	0.8 \\
     ~~~~Simulation size   & 1.1 &	1.6 &	0.1 &	3.4 &	0.8 \\
     ~~~~Hardware trigger efficiency   & 3.3 &	10.6\phz &	0.3 &	3.6 &	1.2 \\
     ~~~~BDT selection efficiency  & 1.0	& 1.5	& 0.1	& 2.0	& 0.6 \\
     ~~~~Luminosity & 0.1 &	0.2 &	0.0 &	0.4 &	0.1 \\
     \midrule
     Branching fraction uncertainty\cite{PDG2024}   & 2.8 &	5.1 &	0.3 &	11.0\phz &	5.0\\
     \bottomrule
    \end{tabular}}
    \label{tab:allsyst}
\end{table}

\section{Results}
\label{sec:results}

The $R_{DX}$ results are:
\begin{align*}
R_{\Dp K^{*0}} &= ( 1.42 \pm 0.23 \pm 0.07 \pm 0.11 ) \times 10^{-6} , \\
R_{D^{*0} \Kp} &= (1.46 \pm 0.30 \pm 0.11 \pm 0.05 ) \times 10^{-6} , \\   
R_{\Dsp \phi}\ \ \  &= ( 4.0\pm 1.3 \pm 0.2 \pm 0.5) \times 10^{-7 } , \\
R_{\Dz \Kp}\ &= ( 9.7 \pm 1.0 \pm 0.4 \pm 0.3 ) \times 10^{-7} , \\
R_{\Dz \pip}\ \,&= ( 4.6 \pm 4.6 \pm 0.4 \pm 0.3 ) \times 10^{-8},
\end{align*}
where the quoted uncertainties are the statistical, systematic, and external effects, respectively. 
For the \decay{\Bc}{\Dz \Kp} mode, due to the increased size of the dataset, the statistical uncertainty is decreased by a factor of two with respect to the previous measurement~\cite{LHCb-PAPER-2016-058}. The decay \decay{\Bcp}{\Dz \pip} remains unobserved. The CL$_s$ method~\cite{CLs,Cowan:2010js} is used to derive an upper limit at 95\% CL of $R_{\Dz \pip}\ < 1.4 \times 10^{-7}$ ($<1.2 \times 10^{-7}$ at 90\% CL).

The $R_{DX}$ results are converted to branching fractions by dividing out the effective fragmentation ratio, $f_c/f_u = (7.50\pm1.79)\times10^{-3}$, yielding
\begin{align*}
\mathcal{B}(\decay{\Bcp}{\Dp K^{*0}}) &= (1.89 \pm 0.31 \pm 0.10 \pm 0.47) \times 10^{-4} , \\
\mathcal{B}(\decay{\Bcp}{\Dstarz \Kp}) &=  (1.94 \pm 0.41 \pm 0.15 \pm 0.47) \times 10^{-4} ,\\  
\mathcal{B}(\decay{\Bcp}{\Dsp \phi})\ \ \ \,&=(5.3 \pm 1.7 \pm 0.2 \pm 1.5) \times 10^{-5} ,\\
\mathcal{B}(\decay{\Bcp}{\Dz \Kp})\ &= (1.30 \pm 0.14 \pm 0.05 \pm 0.31) \times 10^{-4} ,\\
\mathcal{B}(\decay{\Bcp}{\Dz \pip})\ \ & <\ 1.9(1.6) \times 10^{-5} \text{\ at 95\%(90\%) CL,}
\end{align*}
where the first uncertainty is statistical, the second is due to systematic effects, and the third is due to external branching fraction inputs. The effective value of $f_c/f_u$ is obtained by combining the $\sqrt{s}=13\tev$ and $7\tev$ values from \cite{LHCb-PAPER-2019-033} with appropriate weights, and its uncertainty is included in the third uncertainty.

\section{Summary}
A search for $\Bcp$ mesons decaying to a charm and charmless meson pair has been conducted. 
The \decay{\Bcp}{\Dp \Kstarz} decay is observed, and there is strong evidence for the quasi-two-body modes: \decay{\Bcp}{\Dstarz \Kp} and \decay{\Bcp}{\Dsp \phi}. This is consistent with the recent observation of \decay{\Bcp}{D^+_{(s)}h^+h^-} decays\cite{LHCb-PAPER-2025-028}.
The existence of the \decay{\Bcp}{\Dz \Kp} decay is confirmed, halving the statistical uncertainty of a previous result~\cite{LHCb-PAPER-2016-058}. No evidence for the \decay{\Bcp}{\Dz \pip} mode is seen, proving that the \decay{b}{u} amplitude is sub-dominant. A search for direct \CP asymmetry in \decay{\Bcp}{\Dz \Kp} decays is performed, but none is found, as expected in the absence of a significant \decay{b}{u} contribution.
The final state requiring an $s\overline s$ pair in the hadronisation is significantly less likely than those involving light-quark pairs, and the vector meson final states, $\Dstarz\Kp$ and $\Dp\Kstarz$, are $\sim1.5$ times more likely than the \decay{\Bcp}{\Dz\Kp} mode, which is a pattern also seen with \decay{\B^{+,0}}{D^{(*)}K^{(*)}} decays. 
The derived branching fractions are two orders of magnitude higher than the expectation for the $W$-emission and penguin amplitudes alone, and thus corroborate the conclusion that weak annihilation dominates in these \Bcp decays. 


%% file: acknowledgements.tex
\section*{Acknowledgements}
%
%
\noindent We express our gratitude to our colleagues in the CERN
accelerator departments for the excellent performance of the LHC. We
thank the technical and administrative staff at the LHCb
institutes.
We acknowledge support from CERN and from the national agencies:
ARC (Australia);
CAPES, CNPq, FAPERJ and FINEP (Brazil); 
MOST and NSFC (China); 
CNRS/IN2P3 and CEA (France);  
BMFTR, DFG and MPG (Germany);
NKFIH (Hungary);              
INFN (Italy); 
NWO (Netherlands); 
MNiSW and NCN (Poland); 
MEC/IFA (Romania); 
MICIU and AEI (Spain);
SNSF and SER (Switzerland); 
NASU (Ukraine); 
STFC (United Kingdom); 
DOE NP and NSF (USA).
We acknowledge the computing resources that are provided by ARDC (Australia), 
CBPF (Brazil),
CERN, 
IHEP and LZU (China),
IN2P3 (France), 
KIT and DESY (Germany), 
INFN (Italy), 
SURF (Netherlands),
Polish WLCG (Poland),
IFIN-HH (Romania), 
PIC (Spain), CSCS (Switzerland), 
GridPP (United Kingdom),
and NSF (USA).  
We are indebted to the communities behind the multiple open-source
software packages on which we depend.
Individual groups or members have received support from
RTP (Australia), 
FWO Odysseus grant G0ASD25N (Belgium), 
Key Research Program of Frontier Sciences of CAS, CAS PIFI, CAS CCEPP (China); 
Minciencias (Colombia);
EPLANET, Marie Sk\l{}odowska-Curie Actions, ERC and NextGenerationEU (European Union);
A*MIDEX, ANR, IPhU and Labex P2IO, and R\'{e}gion Auvergne-Rh\^{o}ne-Alpes (France);
Alexander-von-Humboldt Foundation (Germany);
ICSC (Italy); 
Severo Ochoa and Mar\'ia de Maeztu Units of Excellence, GVA, XuntaGal, GENCAT, InTalent-Inditex and Prog.~Atracci\'on Talento CM (Spain);
the Leverhulme Trust, the Royal Society and UKRI (United Kingdom).

%% file: Authorship_LHCb-PAPER-2024-035.tex
\centerline
{\large\bf LHCb collaboration}
\begin
{flushleft}
\small
R.~Aaij$^{38}$\lhcborcid{0000-0003-0533-1952},
A.S.W.~Abdelmotteleb$^{57}$\lhcborcid{0000-0001-7905-0542},
C.~Abellan~Beteta$^{51}$\lhcborcid{0009-0009-0869-6798},
F.~Abudin\'en$^{57}$\lhcborcid{0000-0002-6737-3528},
T.~Ackernley$^{61}$\lhcborcid{0000-0002-5951-3498},
A.A.~Adefisoye$^{69}$\lhcborcid{0000-0003-2448-1550},
B.~Adeva$^{47}$\lhcborcid{0000-0001-9756-3712},
M.~Adinolfi$^{55}$\lhcborcid{0000-0002-1326-1264},
P.~Adlarson$^{82}$\lhcborcid{0000-0001-6280-3851},
C.~Agapopoulou$^{14}$\lhcborcid{0000-0002-2368-0147}
% ...

{\footnotesize \it

% ...
$^{14}$Universit{\'e} Paris-Saclay, CNRS/IN2P3, IJCLab, Orsay, France\\
% ...
$^{38}$Nikhef National Institute for Subatomic Physics, Amsterdam, Netherlands\\
% ...
$^{47}$Instituto Galego de F{\'\i}sica de Altas Enerx{\'\i}as (IGFAE), Universidade de Santiago de Compostela, Santiago de Compostela, Spain\\
% ...
$^{51}$Physik-Institut, Universit{\"a}t Z{\"u}rich, Z{\"u}rich, Switzerland\\
% ...
$^{55}$H.H. Wills Physics Laboratory, University of Bristol, Bristol, United Kingdom\\
% ...
$^{57}$Department of Physics, University of Warwick, Coventry, United Kingdom\\
% ...
$^{60}$School of Physics and Astronomy, University of Glasgow, Glasgow, United Kingdom\\
$^{61}$Oliver Lodge Laboratory, University of Liverpool, Liverpool, United Kingdom\\
% ...
$^{69}$Syracuse University, Syracuse, NY, United States\\
% ...
$^{82}$Department of Physics and Astronomy, Uppsala University, Uppsala, Sweden, associated to $^{60}$\\
% ...
}
\end{flushleft}

%% file: LHCb-PAPER-2024-035.bbl
\ifx\mcitethebibliography\mciteundefinedmacro
\PackageError{LHCb.bst}{mciteplus.sty has not been loaded}
{This bibstyle requires the use of the mciteplus package.}\fi
\providecommand{\href}[2]{#2}
\begin{mcitethebibliography}{10}
\mciteSetBstSublistMode{n}
\mciteSetBstMaxWidthForm{subitem}{\alph{mcitesubitemcount})}
\mciteSetBstSublistLabelBeginEnd{\mcitemaxwidthsubitemform\space}
{\relax}{\relax}

\bibitem{LHCb-PAPER-2019-033}
LHCb collaboration, R.~Aaij {\em et~al.},
  \ifthenelse{\boolean{articletitles}}{\emph{{Measurement of the \Bcm
  production fraction and asymmetry in 7 and 13~\tev $pp$ collisions}},
  }{}\href{https://doi.org/10.1103/PhysRevD.100.112006}{Phys.\ ~Rev.\
  \textbf{D100} (2019) 112006},
  \href{http://arxiv.org/abs/1910.13404}{{\normalfont\ttfamily
  arXiv:1910.13404}}\relax
\mciteBstWouldAddEndPuncttrue
\mciteSetBstMidEndSepPunct{\mcitedefaultmidpunct}
{\mcitedefaultendpunct}{\mcitedefaultseppunct}\relax
\EndOfBibitem
\bibitem{LHCb-PAPER-2014-060}
LHCb collaboration, R.~Aaij {\em et~al.},
  \ifthenelse{\boolean{articletitles}}{\emph{{Measurement of the lifetime of
  the \Bcp meson using the \mbox{\decay{\Bc}{\jpsi\pip}} decay mode}},
  }{}\href{https://doi.org/10.1016/j.physletb.2015.01.010}{Phys.\ Lett.\
  \textbf{B742} (2015) 29},
  \href{http://arxiv.org/abs/1411.6899}{{\normalfont\ttfamily
  arXiv:1411.6899}}\relax
\mciteBstWouldAddEndPuncttrue
\mciteSetBstMidEndSepPunct{\mcitedefaultmidpunct}
{\mcitedefaultendpunct}{\mcitedefaultseppunct}\relax
\EndOfBibitem
\bibitem{LHCb-PAPER-2013-044}
LHCb collaboration, R.~Aaij {\em et~al.},
  \ifthenelse{\boolean{articletitles}}{\emph{{Observation of the decay
  \mbox{\decay{\Bc}{\Bs\pip}}}},
  }{}\href{https://doi.org/10.1103/PhysRevLett.111.181801}{Phys.\ ~Rev.\
  ~Lett.\  \textbf{111} (2013) 181801},
  \href{http://arxiv.org/abs/1308.4544}{{\normalfont\ttfamily
  arXiv:1308.4544}}\relax
\mciteBstWouldAddEndPuncttrue
\mciteSetBstMidEndSepPunct{\mcitedefaultmidpunct}
{\mcitedefaultendpunct}{\mcitedefaultseppunct}\relax
\EndOfBibitem
\bibitem{LHCb-PAPER-2016-001}
LHCb collaboration, R.~Aaij {\em et~al.},
  \ifthenelse{\boolean{articletitles}}{\emph{{Search for \Bcp decays to the
  $\proton\antiproton\pip$ final state}},
  }{}\href{https://doi.org/10.1016/j.physletb.2016.05.074}{Phys.\ Lett.\
  \textbf{B759} (2016) 313},
  \href{http://arxiv.org/abs/1603.07037}{{\normalfont\ttfamily
  arXiv:1603.07037}}\relax
\mciteBstWouldAddEndPuncttrue
\mciteSetBstMidEndSepPunct{\mcitedefaultmidpunct}
{\mcitedefaultendpunct}{\mcitedefaultseppunct}\relax
\EndOfBibitem
\bibitem{LHCb-PAPER-2016-058}
LHCb collaboration, R.~Aaij {\em et~al.},
  \ifthenelse{\boolean{articletitles}}{\emph{{Observation of
  \mbox{\decay{\Bc}{\Dz\Kp}} decays}},
  }{}\href{https://doi.org/10.1103/PhysRevLett.118.111803}{Phys.\ ~Rev.\
  ~Lett.\  \textbf{118} (2017) 111803},
  \href{http://arxiv.org/abs/1701.01856}{{\normalfont\ttfamily
  arXiv:1701.01856}}\relax
\mciteBstWouldAddEndPuncttrue
\mciteSetBstMidEndSepPunct{\mcitedefaultmidpunct}
{\mcitedefaultendpunct}{\mcitedefaultseppunct}\relax
\EndOfBibitem
\bibitem{LHCb-PAPER-2025-028}
LHCb collaboration, R.~Aaij {\em et~al.},
  \ifthenelse{\boolean{articletitles}}{\emph{{Observation of $\Bc \to D h^+
  h^-$ decays}}, }{}\href{https://doi.org/10.1103/fsj9-89jt}{{Phys.\ ~Rev.\
  ~Lett.\ } \textbf{136} (2026) 021804},
  \href{http://arxiv.org/abs/2509.15873}{{\normalfont\ttfamily
  arXiv:2509.15873}}\relax
\mciteBstWouldAddEndPuncttrue
\mciteSetBstMidEndSepPunct{\mcitedefaultmidpunct}
{\mcitedefaultendpunct}{\mcitedefaultseppunct}\relax
\EndOfBibitem
\bibitem{PDG2024}
Particle Data Group, S.~Navas {\em et~al.},
  \ifthenelse{\boolean{articletitles}}{\emph{{\href{http://pdg.lbl.gov/}{Review
  of particle physics}}},
  }{}\href{https://doi.org/10.1103/PhysRevD.110.030001}{Phys.\ Rev.\
  \textbf{D110} (2024) 030001}\relax
\mciteBstWouldAddEndPuncttrue
\mciteSetBstMidEndSepPunct{\mcitedefaultmidpunct}
{\mcitedefaultendpunct}{\mcitedefaultseppunct}\relax
\EndOfBibitem
\bibitem{zhang2009branching}
J.~Zhang and X.-Q. Yu, \ifthenelse{\boolean{articletitles}}{\emph{Branching
  ratio and {CP} violation of ${B}_c^+ \to {D}^0 {K}^+$ decays in the
  perturbative {QCD} approach},
  }{}\href{https://doi.org/10.1140/epjc/s10052-009-1112-1}{Eur.\ Phys.\ J.\
  \textbf{C63} (2009) 435–442},
  \href{http://arxiv.org/abs/0905.0945}{{\normalfont\ttfamily
  arXiv:0905.0945}}\relax
\mciteBstWouldAddEndPuncttrue
\mciteSetBstMidEndSepPunct{\mcitedefaultmidpunct}
{\mcitedefaultendpunct}{\mcitedefaultseppunct}\relax
\EndOfBibitem
\bibitem{LHCb-DP-2008-001}
LHCb collaboration, A.~A. Alves~Jr.\ {\em et~al.},
  \ifthenelse{\boolean{articletitles}}{\emph{{The \lhcb detector at the LHC}},
  }{}\href{https://doi.org/10.1088/1748-0221/3/08/S08005}{JINST \textbf{3}
  (2008) S08005} LHCb-DP-2008-001\relax
\mciteBstWouldAddEndPuncttrue
\mciteSetBstMidEndSepPunct{\mcitedefaultmidpunct}
{\mcitedefaultendpunct}{\mcitedefaultseppunct}\relax
\EndOfBibitem
\bibitem{LHCb-DP-2014-002}
LHCb collaboration, R.~Aaij {\em et~al.},
  \ifthenelse{\boolean{articletitles}}{\emph{{LHCb detector performance}},
  }{}\href{https://doi.org/10.1142/S0217751X15300227}{Int.\ J.\ Mod.\ Phys.\
  \textbf{A30} (2015) 1530022},
  \href{http://arxiv.org/abs/1412.6352}{{\normalfont\ttfamily
  arXiv:1412.6352}}\relax
\mciteBstWouldAddEndPuncttrue
\mciteSetBstMidEndSepPunct{\mcitedefaultmidpunct}
{\mcitedefaultendpunct}{\mcitedefaultseppunct}\relax
\EndOfBibitem
\bibitem{LHCb-DP-2014-001}
R.~Aaij {\em et~al.}, \ifthenelse{\boolean{articletitles}}{\emph{{Performance
  of the LHCb Vertex Locator}},
  }{}\href{https://doi.org/10.1088/1748-0221/9/09/P09007}{JINST \textbf{9}
  (2014) P09007}, \href{http://arxiv.org/abs/1405.7808}{{\normalfont\ttfamily
  arXiv:1405.7808}}\relax
\mciteBstWouldAddEndPuncttrue
\mciteSetBstMidEndSepPunct{\mcitedefaultmidpunct}
{\mcitedefaultendpunct}{\mcitedefaultseppunct}\relax
\EndOfBibitem
\bibitem{LHCb-DP-2013-003}
R.~Arink {\em et~al.}, \ifthenelse{\boolean{articletitles}}{\emph{{Performance
  of the LHCb Outer Tracker}},
  }{}\href{https://doi.org/10.1088/1748-0221/9/01/P01002}{JINST \textbf{9}
  (2014) P01002}, \href{http://arxiv.org/abs/1311.3893}{{\normalfont\ttfamily
  arXiv:1311.3893}}\relax
\mciteBstWouldAddEndPuncttrue
\mciteSetBstMidEndSepPunct{\mcitedefaultmidpunct}
{\mcitedefaultendpunct}{\mcitedefaultseppunct}\relax
\EndOfBibitem
\bibitem{LHCb-DP-2017-001}
P.~d'Argent {\em et~al.}, \ifthenelse{\boolean{articletitles}}{\emph{{Improved
  performance of the LHCb Outer Tracker in LHC Run 2}},
  }{}\href{https://doi.org/10.1088/1748-0221/12/11/P11016}{JINST \textbf{12}
  (2017) P11016}, \href{http://arxiv.org/abs/1708.00819}{{\normalfont\ttfamily
  arXiv:1708.00819}}\relax
\mciteBstWouldAddEndPuncttrue
\mciteSetBstMidEndSepPunct{\mcitedefaultmidpunct}
{\mcitedefaultendpunct}{\mcitedefaultseppunct}\relax
\EndOfBibitem
\bibitem{LHCb-DP-2012-003}
M.~Adinolfi {\em et~al.},
  \ifthenelse{\boolean{articletitles}}{\emph{{Performance of the \lhcb RICH
  detector at the LHC}},
  }{}\href{https://doi.org/10.1140/epjc/s10052-013-2431-9}{Eur.\ Phys.\ J.\
  \textbf{C73} (2013) 2431},
  \href{http://arxiv.org/abs/1211.6759}{{\normalfont\ttfamily
  arXiv:1211.6759}}\relax
\mciteBstWouldAddEndPuncttrue
\mciteSetBstMidEndSepPunct{\mcitedefaultmidpunct}
{\mcitedefaultendpunct}{\mcitedefaultseppunct}\relax
\EndOfBibitem
\bibitem{LHCb-DP-2012-002}
A.~A. Alves~Jr.\ {\em et~al.},
  \ifthenelse{\boolean{articletitles}}{\emph{{Performance of the LHCb muon
  system}}, }{}\href{https://doi.org/10.1088/1748-0221/8/02/P02022}{JINST
  \textbf{8} (2013) P02022},
  \href{http://arxiv.org/abs/1211.1346}{{\normalfont\ttfamily
  arXiv:1211.1346}}\relax
\mciteBstWouldAddEndPuncttrue
\mciteSetBstMidEndSepPunct{\mcitedefaultmidpunct}
{\mcitedefaultendpunct}{\mcitedefaultseppunct}\relax
\EndOfBibitem
\bibitem{LHCb-DP-2012-004}
R.~Aaij {\em et~al.}, \ifthenelse{\boolean{articletitles}}{\emph{{The \lhcb
  trigger and its performance in 2011}},
  }{}\href{https://doi.org/10.1088/1748-0221/8/04/P04022}{JINST \textbf{8}
  (2013) P04022}, \href{http://arxiv.org/abs/1211.3055}{{\normalfont\ttfamily
  arXiv:1211.3055}}\relax
\mciteBstWouldAddEndPuncttrue
\mciteSetBstMidEndSepPunct{\mcitedefaultmidpunct}
{\mcitedefaultendpunct}{\mcitedefaultseppunct}\relax
\EndOfBibitem
\bibitem{Sjostrand:2007gs}
T.~Sj\"{o}strand, S.~Mrenna, and P.~Skands,
  \ifthenelse{\boolean{articletitles}}{\emph{{A brief introduction to PYTHIA
  8.1}}, }{}\href{https://doi.org/10.1016/j.cpc.2008.01.036}{Comput.\ Phys.\
  Commun.\  \textbf{178} (2008) 852},
  \href{http://arxiv.org/abs/0710.3820}{{\normalfont\ttfamily
  arXiv:0710.3820}}\relax
\mciteBstWouldAddEndPuncttrue
\mciteSetBstMidEndSepPunct{\mcitedefaultmidpunct}
{\mcitedefaultendpunct}{\mcitedefaultseppunct}\relax
\EndOfBibitem
\bibitem{Sjostrand:2006za}
T.~Sj\"{o}strand, S.~Mrenna, and P.~Skands,
  \ifthenelse{\boolean{articletitles}}{\emph{{PYTHIA 6.4 physics and manual}},
  }{}\href{https://doi.org/10.1088/1126-6708/2006/05/026}{JHEP \textbf{05}
  (2006) 026}, \href{http://arxiv.org/abs/hep-ph/0603175}{{\normalfont\ttfamily
  arXiv:hep-ph/0603175}}\relax
\mciteBstWouldAddEndPuncttrue
\mciteSetBstMidEndSepPunct{\mcitedefaultmidpunct}
{\mcitedefaultendpunct}{\mcitedefaultseppunct}\relax
\EndOfBibitem
\bibitem{LHCb-PROC-2010-056}
I.~Belyaev {\em et~al.}, \ifthenelse{\boolean{articletitles}}{\emph{{Handling
  of the generation of primary events in Gauss, the LHCb simulation
  framework}}, }{}\href{https://doi.org/10.1088/1742-6596/331/3/032047}{J.\
  Phys.\ Conf.\ Ser.\  \textbf{331} (2011) 032047}\relax
\mciteBstWouldAddEndPuncttrue
\mciteSetBstMidEndSepPunct{\mcitedefaultmidpunct}
{\mcitedefaultendpunct}{\mcitedefaultseppunct}\relax
\EndOfBibitem
\bibitem{Chang_2004}
C.-H. Chang, C.~Driouichi, P.~Eerola, and X.-G. Wu,
  \ifthenelse{\boolean{articletitles}}{\emph{{BCVEGPY: an event generator for
  hadronic production of the $B_c$ meson}},
  }{}\href{https://doi.org/10.1016/j.cpc.2004.02.005}{Comput.\ Phys.\ Commun.\
  \textbf{159} (2004) 192–224}\relax
\mciteBstWouldAddEndPuncttrue
\mciteSetBstMidEndSepPunct{\mcitedefaultmidpunct}
{\mcitedefaultendpunct}{\mcitedefaultseppunct}\relax
\EndOfBibitem
\bibitem{Chang_2015}
C.-H. Chang, X.-Y. Wang, and X.-G. Wu,
  \ifthenelse{\boolean{articletitles}}{\emph{{BCVEGPY2.2: A newly upgraded
  version for hadronic production of the meson $B_c$ and its excited states}},
  }{}\href{https://doi.org/10.1016/j.cpc.2015.07.015}{Comput.\ Phys.\ Commun.\
  \textbf{197} (2015) 335–338}\relax
\mciteBstWouldAddEndPuncttrue
\mciteSetBstMidEndSepPunct{\mcitedefaultmidpunct}
{\mcitedefaultendpunct}{\mcitedefaultseppunct}\relax
\EndOfBibitem
\bibitem{Lange:2001uf}
D.~J. Lange, \ifthenelse{\boolean{articletitles}}{\emph{{The EvtGen particle
  decay simulation package}},
  }{}\href{https://doi.org/10.1016/S0168-9002(01)00089-4}{Nucl.\ Instrum.\
  Meth.\  \textbf{A462} (2001) 152}\relax
\mciteBstWouldAddEndPuncttrue
\mciteSetBstMidEndSepPunct{\mcitedefaultmidpunct}
{\mcitedefaultendpunct}{\mcitedefaultseppunct}\relax
\EndOfBibitem
\bibitem{davidson2015photos}
N.~Davidson, T.~Przedzinski, and Z.~Was,
  \ifthenelse{\boolean{articletitles}}{\emph{{PHOTOS interface in C++:
  Technical and physics documentation}},
  }{}\href{https://doi.org/https://doi.org/10.1016/j.cpc.2015.09.013}{Comp.\
  Phys.\ Comm.\  \textbf{199} (2016) 86},
  \href{http://arxiv.org/abs/1011.0937}{{\normalfont\ttfamily
  arXiv:1011.0937}}\relax
\mciteBstWouldAddEndPuncttrue
\mciteSetBstMidEndSepPunct{\mcitedefaultmidpunct}
{\mcitedefaultendpunct}{\mcitedefaultseppunct}\relax
\EndOfBibitem
\bibitem{Allison:2006ve}
Geant4 collaboration, J.~Allison {\em et~al.},
  \ifthenelse{\boolean{articletitles}}{\emph{{Geant4 developments and
  applications}}, }{}\href{https://doi.org/10.1109/TNS.2006.869826}{IEEE
  Trans.\ Nucl.\ Sci.\  \textbf{53} (2006) 270}\relax
\mciteBstWouldAddEndPuncttrue
\mciteSetBstMidEndSepPunct{\mcitedefaultmidpunct}
{\mcitedefaultendpunct}{\mcitedefaultseppunct}\relax
\EndOfBibitem
\bibitem{Agostinelli:2002hh}
Geant4 collaboration, S.~Agostinelli {\em et~al.},
  \ifthenelse{\boolean{articletitles}}{\emph{{Geant4: A simulation toolkit}},
  }{}\href{https://doi.org/10.1016/S0168-9002(03)01368-8}{Nucl.\ Instrum.\
  Meth.\  \textbf{A506} (2003) 250}\relax
\mciteBstWouldAddEndPuncttrue
\mciteSetBstMidEndSepPunct{\mcitedefaultmidpunct}
{\mcitedefaultendpunct}{\mcitedefaultseppunct}\relax
\EndOfBibitem
\bibitem{LHCb-PROC-2011-006}
M.~Clemencic {\em et~al.}, \ifthenelse{\boolean{articletitles}}{\emph{{The
  \lhcb simulation application, Gauss: Design, evolution and experience}},
  }{}\href{https://doi.org/10.1088/1742-6596/331/3/032023}{J.\ Phys.\ Conf.\
  Ser.\  \textbf{331} (2011) 032023}\relax
\mciteBstWouldAddEndPuncttrue
\mciteSetBstMidEndSepPunct{\mcitedefaultmidpunct}
{\mcitedefaultendpunct}{\mcitedefaultseppunct}\relax
\EndOfBibitem
\bibitem{LHCb-DP-2018-004}
D.~M{\"u}ller, M.~Clemencic, G.~Corti, and M.~Gersabeck,
  \ifthenelse{\boolean{articletitles}}{\emph{{ReDecay: A novel approach to
  speed up the simulation at LHCb}},
  }{}\href{https://doi.org/10.1140/epjc/s10052-018-6469-6}{Eur.\ Phys.\ J.\
  \textbf{C78} (2018) 1009},
  \href{http://arxiv.org/abs/1810.10362}{{\normalfont\ttfamily
  arXiv:1810.10362}}\relax
\mciteBstWouldAddEndPuncttrue
\mciteSetBstMidEndSepPunct{\mcitedefaultmidpunct}
{\mcitedefaultendpunct}{\mcitedefaultseppunct}\relax
\EndOfBibitem
\bibitem{Cowan:2016tnm}
G.~A. Cowan, D.~C. Craik, and M.~D. Needham,
  \ifthenelse{\boolean{articletitles}}{\emph{{RapidSim: an application for the
  fast simulation of heavy-quark hadron decays}},
  }{}\href{https://doi.org/10.1016/j.cpc.2017.01.029}{Comput.\ Phys.\ Commun.\
  \textbf{214} (2017) 239},
  \href{http://arxiv.org/abs/1612.07489}{{\normalfont\ttfamily
  arXiv:1612.07489}}\relax
\mciteBstWouldAddEndPuncttrue
\mciteSetBstMidEndSepPunct{\mcitedefaultmidpunct}
{\mcitedefaultendpunct}{\mcitedefaultseppunct}\relax
\EndOfBibitem
\bibitem{LHCb-PUB-2016-021}
L.~Anderlini {\em et~al.}, \ifthenelse{\boolean{articletitles}}{\emph{{The
  PIDCalib package}}, }{}
  \href{https://cds.cern.ch/search?p=LHCb-PUB-2016-021&f=reportnumber&action_search=Search&c=LHCb+Notes}
  {LHCb-PUB-2016-021}, 2016\relax
\mciteBstWouldAddEndPuncttrue
\mciteSetBstMidEndSepPunct{\mcitedefaultmidpunct}
{\mcitedefaultendpunct}{\mcitedefaultseppunct}\relax
\EndOfBibitem
\bibitem{LHCb-DP-2018-001}
R.~Aaij {\em et~al.}, \ifthenelse{\boolean{articletitles}}{\emph{{Selection and
  processing of calibration samples to measure the particle identification
  performance of the LHCb experiment in Run 2}},
  }{}\href{https://doi.org/10.1140/epjti/s40485-019-0050-z}{Eur.\ Phys.\ J.\
  Tech.\ Instr.\  \textbf{6} (2019) 1},
  \href{http://arxiv.org/abs/1803.00824}{{\normalfont\ttfamily
  arXiv:1803.00824}}\relax
\mciteBstWouldAddEndPuncttrue
\mciteSetBstMidEndSepPunct{\mcitedefaultmidpunct}
{\mcitedefaultendpunct}{\mcitedefaultseppunct}\relax
\EndOfBibitem
\bibitem{LHCb-DP-2019-001}
R.~Aaij {\em et~al.}, \ifthenelse{\boolean{articletitles}}{\emph{{Design and
  performance of the LHCb trigger and full real-time reconstruction in Run 2 of
  the LHC}}, }{}\href{https://doi.org/10.1088/1748-0221/14/04/P04013}{JINST
  \textbf{14} (2019) P04013},
  \href{http://arxiv.org/abs/1812.10790}{{\normalfont\ttfamily
  arXiv:1812.10790}}\relax
\mciteBstWouldAddEndPuncttrue
\mciteSetBstMidEndSepPunct{\mcitedefaultmidpunct}
{\mcitedefaultendpunct}{\mcitedefaultseppunct}\relax
\EndOfBibitem
\bibitem{BBDT}
V.~V. Gligorov and M.~Williams,
  \ifthenelse{\boolean{articletitles}}{\emph{{Efficient, reliable and fast
  high-level triggering using a bonsai boosted decision tree}},
  }{}\href{https://doi.org/10.1088/1748-0221/8/02/P02013}{JINST \textbf{8}
  (2013) P02013}, \href{http://arxiv.org/abs/1210.6861}{{\normalfont\ttfamily
  arXiv:1210.6861}}\relax
\mciteBstWouldAddEndPuncttrue
\mciteSetBstMidEndSepPunct{\mcitedefaultmidpunct}
{\mcitedefaultendpunct}{\mcitedefaultseppunct}\relax
\EndOfBibitem
\bibitem{Hulsbergen:2005pu}
W.~D. Hulsbergen, \ifthenelse{\boolean{articletitles}}{\emph{{Decay chain
  fitting with a Kalman filter}},
  }{}\href{https://doi.org/10.1016/j.nima.2005.06.078}{Nucl.\ Instrum.\ Meth.\
  \textbf{A552} (2005) 566},
  \href{http://arxiv.org/abs/physics/0503191}{{\normalfont\ttfamily
  arXiv:physics/0503191}}\relax
\mciteBstWouldAddEndPuncttrue
\mciteSetBstMidEndSepPunct{\mcitedefaultmidpunct}
{\mcitedefaultendpunct}{\mcitedefaultseppunct}\relax
\EndOfBibitem
\bibitem{Roe:2004na}
B.~P. Roe {\em et~al.}, \ifthenelse{\boolean{articletitles}}{\emph{{Boosted
  decision trees, an alternative to artificial neural networks}},
  }{}\href{https://doi.org/10.1016/j.nima.2004.12.018}{Nucl.\ Instrum.\ Meth.\
  \textbf{A543} (2005) 577},
  \href{http://arxiv.org/abs/physics/0408124}{{\normalfont\ttfamily
  arXiv:physics/0408124}}\relax
\mciteBstWouldAddEndPuncttrue
\mciteSetBstMidEndSepPunct{\mcitedefaultmidpunct}
{\mcitedefaultendpunct}{\mcitedefaultseppunct}\relax
\EndOfBibitem
\bibitem{Skwarnicki:1986xj}
T.~Skwarnicki, {\em {A study of the radiative cascade transitions between the
  Upsilon-prime and Upsilon resonances}}, PhD thesis, Institute of Nuclear
  Physics, Krakow, 1986,
  {\href{http://inspirehep.net/record/230779/}{DESY-F31-86-02}}\relax
\mciteBstWouldAddEndPuncttrue
\mciteSetBstMidEndSepPunct{\mcitedefaultmidpunct}
{\mcitedefaultendpunct}{\mcitedefaultseppunct}\relax
\EndOfBibitem
\bibitem{LHCb-PAPER-2020-036}
LHCb collaboration, R.~Aaij {\em et~al.},
  \ifthenelse{\boolean{articletitles}}{\emph{{Measurement of \CP observables in
  $\Bpm \to D^{(*)} \Kpm$ and $\Bpm \to D^{(*)} \pipm $ decays using two-body
  $D$ final states}}, }{}\href{https://doi.org/10.1007/JHEP04(2021)081}{JHEP
  \textbf{04} (2021) 081},
  \href{http://arxiv.org/abs/2012.09903}{{\normalfont\ttfamily
  arXiv:2012.09903}}\relax
\mciteBstWouldAddEndPuncttrue
\mciteSetBstMidEndSepPunct{\mcitedefaultmidpunct}
{\mcitedefaultendpunct}{\mcitedefaultseppunct}\relax
\EndOfBibitem
\bibitem{Wilks:1938dza}
S.~S. Wilks, \ifthenelse{\boolean{articletitles}}{\emph{{The large-sample
  distribution of the likelihood ratio for testing composite hypotheses}},
  }{}\href{https://doi.org/10.1214/aoms/1177732360}{Ann.\ Math.\ Stat.\
  \textbf{9} (1938) 60}\relax
\mciteBstWouldAddEndPuncttrue
\mciteSetBstMidEndSepPunct{\mcitedefaultmidpunct}
{\mcitedefaultendpunct}{\mcitedefaultseppunct}\relax
\EndOfBibitem
\bibitem{LHCb-PUB-2014-039}
S.~Tolk, J.~Albrecht, F.~Dettori, and A.~Pellegrino,
  \ifthenelse{\boolean{articletitles}}{\emph{{Data driven trigger efficiency
  determination at LHCb}}, }{}
  \href{https://cds.cern.ch/search?p=LHCb-PUB-2014-039&f=reportnumber&action_search=Search&c=LHCb+Notes}
  {LHCb-PUB-2014-039}, 2014\relax
\mciteBstWouldAddEndPuncttrue
\mciteSetBstMidEndSepPunct{\mcitedefaultmidpunct}
{\mcitedefaultendpunct}{\mcitedefaultseppunct}\relax
\EndOfBibitem
\bibitem{CLs}
A.~L. Read, \ifthenelse{\boolean{articletitles}}{\emph{{Presentation of search
  results: The CL$_{\rm s}$ technique}},
  }{}\href{https://doi.org/10.1088/0954-3899/28/10/313}{J.\ Phys.\
  \textbf{G28} (2002) 2693}\relax
\mciteBstWouldAddEndPuncttrue
\mciteSetBstMidEndSepPunct{\mcitedefaultmidpunct}
{\mcitedefaultendpunct}{\mcitedefaultseppunct}\relax
\EndOfBibitem
\bibitem{Cowan:2010js}
G.~Cowan, K.~Cranmer, E.~Gross, and O.~Vitells,
  \ifthenelse{\boolean{articletitles}}{\emph{{Asymptotic formulae for
  likelihood-based tests of new physics}},
  }{}\href{https://doi.org/10.1140/epjc/s10052-011-1554-0}{Eur.\ Phys.\ J.\
  \textbf{C71} (2011) 1554}, Erratum
  \href{https://doi.org/10.1140/epjc/s10052-013-2501-z}{ibid.\   \textbf{C73}
  (2013) 2501}, \href{http://arxiv.org/abs/1007.1727}{{\normalfont\ttfamily
  arXiv:1007.1727}}\relax
\mciteBstWouldAddEndPuncttrue
\mciteSetBstMidEndSepPunct{\mcitedefaultmidpunct}
{\mcitedefaultendpunct}{\mcitedefaultseppunct}\relax
\EndOfBibitem
\end{mcitethebibliography}
